\documentclass[pra,a4paper,nofootinbib,reprint,showpacs]{revtex4-1}
\usepackage{amsmath,amssymb,amsfonts}
\usepackage{graphicx}
\usepackage{txfonts}
\usepackage{color}

\usepackage[unicode,breaklinks]{hyperref}
\hypersetup{
    unicode=true,
    a4paper=true,
    plainpages=false,
    pdftitle={Non-equilibrium dynamics of a driven Bose-Einstein
condensate at finite temperature},
    pdfauthor={T. P. Billam, P. Mason, and S. A. Gardiner},
    pdfsubject={Non-equilibrium dynamics of a driven Bose-Einstein
condensate at finite temperature},
    colorlinks=true,
    linkcolor=blue,
    citecolor=blue,
    filecolor=black,
    urlcolor=blue
}
\newcommand{\eqnrefp}[1]{{[Eq.~(\ref{#1})]}}
\newcommand{\eqnsrefp}[2]{{[Eqs.~(\ref{#1}) and (\ref{#2})]}}
\newcommand{\eqnreft}[1]{{Eq.~(\ref{#1})}}

\newcommand{\eqnsreft}[2]{{Eqs.~(\ref{#1}) and (\ref{#2})}}
\newcommand{\figreft}[2]{Fig.~\ref{#1}#2}

\newcommand{\figreftfull}[2]{Figure~\ref{#1}#2}
\newcommand{\figrefp}[2]{[Fig.~\ref{#1}#2]}

\newcommand{\secreft}[1]{{Section~\ref{#1}}}

\newcommand{\x}{\theta}
\newcommand{\fo}{\hat{\Psi}(\x)}
\newcommand{\foh}{\hat{\Psi}^\dagger(\x)}
\newcommand{\g}{C}
\newcommand{\avg}[1]{\left\langle #1 \right\rangle}
\newcommand{\twovec}[2]{\left( \begin{array}{c} #1 \\ #2 \end{array} \right)}

\newcommand{\is}{g_{T}}
\newcommand{\vx}{\mathbf{r}}
\newcommand{\vy}{\mathbf{r}^\prime}
\newcommand{\vz}{\mathbf{r}^{\prime\prime}}

\newcommand{\fpsi}{\hat{\Psi}(\vx)}
\newcommand{\fpsid}{\hat{\Psi}^\dagger(\vx)}

\newcommand{\cannil}{\hat{a}_c(t)}

\newcommand{\dfpsi}{\delta\hat{\Psi}(\vx,t)}

\newcommand{\com}[2]{\left[ #1 , #2 \right]}

\newcommand{\bdgvector}[1]{\left( \begin{array}{c} \tilde{\Lambda}(#1) \\ \tilde{\Lambda}^\dagger (#1) \end{array} \right)}

\newcommand{\ltw}[1]{\tilde{\Lambda} (#1)}
\newcommand{\ltwd}[1]{\tilde{\Lambda}^\dagger (#1)}
\newcommand{\ltwpair}[1]{\left( \begin{array}{c} \ltw{#1} \\ \ltwd{#1} \end{array} \right)}
\newcommand{\uvpair}[1]{\left( \begin{array}{c} u_k\left(#1\right) \\ v_k\left(#1\right) \end{array} \right)}
\newcommand{\uvstarpair}[1]{\left( \begin{array}{c} v_k^\ast\left(#1\right) \\ u_k^\ast\left(#1\right) \end{array} \right)}
\newcommand{\qcreate}{\tilde{b}_k^\dagger}
\newcommand{\qannil}{\tilde{b}_k}

\begin{document}

\title{Second-order, number-conserving description of non-equilibrium dynamics in finite-temperature Bose-Einstein
condensates}

\author{T. P. Billam}
\email{thomas.billam@otago.ac.nz}
\affiliation{Jack Dodd Centre for Quantum Technology, Department of Physics,
University of Otago, Dunedin 9016, New Zealand}
\author{P. Mason}
\affiliation{Joint Quantum Centre (JQC) Durham-Newcastle, Department of Physics, Durham University, Durham, DH1 3LE, UK}
\author{S. A. Gardiner}
\affiliation{Joint Quantum Centre (JQC) Durham-Newcastle, Department of Physics, Durham University, Durham, DH1 3LE, UK}

\date{\today}

\pacs{
03.75.Kk     
67.85.De     
05.30.Jp     
}

\begin{abstract}
While the Gross--Pitaevskii equation is well-established as the canonical
dynamical description of atomic Bose-Einstein condensates (BECs) at zero-temperature,
describing the dynamics of BECs at finite temperatures
remains a difficult theoretical problem, particularly when considering
low-temperature, non-equilibrium systems in which depletion of the condensate
occurs dynamically as a result of external driving.  In this paper, we describe
a fully time-dependent numerical implementation of a second-order,
number-conserving description of finite-temperature BEC dynamics.  This
description consists of equations of motion describing the coupled dynamics of
the condensate and non-condensate fractions in a self-consistent manner, and is
ideally suited for the study of low-temperature, non-equilibrium, driven
systems. The $\delta$-kicked-rotor BEC provides a prototypical example of such
a system, and we demonstrate the efficacy of our numerical implementation by
investigating its dynamics at finite temperature.  We demonstrate that the
qualitative features of the system dynamics at zero temperature are generally
preserved at finite temperatures, and predict a quantitative finite-temperature
shift of resonance frequencies which would be relevant for, and could be
verified by, future experiments.
\end{abstract}

\maketitle

\section{Introduction\label{section:nc_intro}}

Even at zero temperature, typical atomic Bose-Einstein condensates (BECs)
contain a small non-condensate fraction due to inter-atomic interactions. In
real experiments, which necessarily take place at finite temperatures and
may involve dynamical depletion of the condensate \cite{gardiner_jmo_2002} ---
for instance, due to non-equilibrium dynamics induced by changes in applied
external fields, a non-negligible non-condensate fraction often arises.  A
full understanding of the dynamics of such systems requires a fully-dynamical,
finite-temperature theoretical description which goes beyond the mean-field,
zero-temperature Gross--Pitaevskii equation (GPE).  Due to the complexity of
such descriptions, the non-equilibrium dynamics of atomic BECs in the presence
of a significant non-condensate fraction (whether thermal or dynamical in
origin) remains a largely open problem \cite{proukakis_jackson_jpb_2008,
gasenzer_pawlowski_plb_2008}. 

In systems where the non-condensate fraction is primarily thermal in origin, a
variety of theoretical descriptions have been developed and successfully
applied to non-equilibrium atomic BEC dynamics.  These include
\textit{symmetry-breaking} descriptions, which are based on a perturbative
expansion about a mean field; for example, the Hartree--Fock--Bogoliubov--Popov
(HFBP) description \cite{hutchinson_etal_prl_1997, dodd_etal_pra_1998,
reidl_etal_pra_1999, giorgini_pra_1998}, and the Zaremba--Nikuni--Griffin (ZNG)
description \cite{zaremba_etal_jltp_1999, griffin_etal_2009,
jackson_zaremba_prl_2001, jackson_zaremba_prl_2002, williams_etal_prl_2002,
jackson_etal_pra_2009, jackson_etal_pra_2007}.  Other successful descriptions
have been obtained in the context of \textit{$c$-field} methods, reviewed in
\cite{blakie_etal_ap_2008}, which describe the highly occupied modes of the
system as a classical field.  These descriptions include the projected
Gross--Pitaevskii equation (PGPE) \cite{davis_etal_prl_2001,
davis_blakie_prl_2006, simula_blakie_prl_2006}, the stochastic projected GPE (SPGPE)
\cite{gardiner_etal_jpb_2002, bradley_etal_pra_2008} and stochastic GPE (SGPE)
\cite{stoof_prl_1997, proukakis_jackson_jpb_2008, cockburn_proukakis_lp_2009,
cockburn_etal_prl_2010, weiler_etal_nature_2008, neely_etal_prl_2010,
fialko_etal_prl_2012}, and the truncated Wigner PGPE
\cite{lobo_etal_prl_2004, norrie_etal_pra_2006, scott_etal_prl_2008,
martin_ruostekoski_njp_2012}.

However, the non-equilibrium dynamics of systems in which a low-temperature BEC is driven by an
applied external field leading to \textit{dynamical} depletion of the condensate have not been as widely investigated. 
The prime example of such systems are atomic BEC analogs of generic
quantum chaotic systems; e.g., the kicked accelerator \cite{ma_etal_prl_2004,
schlunk_etal_prl_2003b}, kicked harmonic oscillator
\cite{billam_gardiner_pra_2009, gardiner_etal_prl_1997,
gardiner_etal_pra_2000}, and kicked rotor \cite{zhang_etal_prl_2004,
liu_etal_pra_2006, shepelyansky_prl_1993, reslen_etal_pra_2008,
monteiro_etal_prl_2009, billam_gardiner_njp_2012}.  These systems offer an
excellent test bed for exploring generic issues of quantum chaos
\cite{schlunk_etal_prl_2003,shepelyansky_prl_1993}, quantum superposition
\cite{weiss_teichmann_prl_2008}, quantum resonances
\cite{billam_gardiner_pra_2009, reslen_etal_pra_2008, monteiro_etal_prl_2009},
dynamical instability and dynamical depletion \cite{gardiner_etal_pra_2000,
monteiro_etal_prl_2009, zhang_etal_prl_2004, reslen_etal_pra_2008}, and even
entropy, thermalization and integrability \cite{polkovkinov_etal_rmp_2011,
yukalov_lpl_2011, yurovsky_olshanii_prl_2011}.  
A particular issue in the description of these systems is the occurence of
nonlinear quantum resonances \cite{monteiro_etal_prl_2009}: values of the
system parameters at which the system resonantly absorbs energy from the
driving. Such resonances are an extremely generic feature of these systems, and
are sensitive to the value of the nonlinearity (product of $s$-wave scattering
length $a_s$ and condensate occupation $N_c$). Consequently, one expects the
dynamical depletion of the condensate caused by the driving to rapidly and
fundamentally alter the subsequent dynamics of the system close to these
resonances, compared to the predictions of the Gross--Pitaevskii equation
\cite{gardiner_jmo_2002}.

The rapid and severe back-action of condensate depletion on the non-equilibrium
dynamics of these systems presents considerable challenges for the
above-mentioned finite-temperature descriptions. In particular, the PGPE and
SPGPE are fundamentally restricted to a high-temperature regime, and lack a
quantum treatment of the pair-excitation processes which drive condensate
depletion.  In principle the truncated-Wigner PGPE may be used to model
non-equilibrium dynamics of driven systems at low (and zero) temperatures
\cite{martin_ruostekoski_finess_2011}. However, in order to go beyond a
close-to-equilibrium approximation in which the condensate remains in its
initial state, one must obtain the dynamics of the condensate within such a
treatment from an ensemble average of individual GPE trajectories. While such a
treatment is entirely possible, and can for example be conducted by determining
the single-particle density matrix in a number-conserving fashion
\cite{wright_etal_pra_2011}, the issues of spurious thermalization and
inaccurate long-time dynamics in the truncated Wigner method (see, e.g., Refs.
\cite{blakie_etal_ap_2008, sinatra_etal_jpb_2002}) would necessitate a very
careful approach when applying such a treatment to the non-equilibrium dynamics
of driven systems. 

In order to comprehensively describe such systems one must thus
self-consistently capture the back action of the increasing population of, and
the subsequent dynamics in, low-lying non-condensate excitations on the
condensate.  To achieve this, a description which consistently divides the
system into well-defined condensate and non-condensate fractions at all times
would appear to be necessary (as opposed to a $c$-field description in which
the condensate fraction must be extracted by subsequent ensemble averaging).
In that they comprise a set of coupled equations for the condensate and the
non-condensate components, symmetry-breaking descriptions would thus appear to
offer a more appropriate treatment of low-temperature driven systems. However,
all such dynamical descriptions which have been applied to BEC dynamics to date
have contained an approximate treatment of pair-excitation processes that
neglects the anomalous average; while suitable at higher temperatures, this
leads to descriptions which incompletely model the phonon character of
elementary excitations at low temperatures \cite{allen_etal_finess_2011}.
Furthermore, the assumption of a symmetry-broken condensate state, and hence
overall number-non-conservation, in any symmetry-breaking method presents an
additional potential problem; since the prime contribution to the departure
from GPE dynamics in low-temperature driven systems results from, and can be
extremely sensitive to, changes in the number of condensate particles,
conservation of the total atom number appears to be necessary in order to
obtain a fully self-consistent description of such systems as realized in
atomic BEC experiments (in which the total atom number is finite and fixed)
\cite{gardiner_morgan_pra_2007}. In a formal sense, coupled equations of motion
for the condensate and non-condensate within a symmetry-breaking treatment do
not contain any terms to explicitly preserve the orthogonality of the
condensate and non-condensate modes (see Section~\ref{section:nc_motivation});
in the context of low-temperature driven systems, this means an ambiguity in
interpreting the solutions of those equations for anything other than short
times.

In view of the above considerations number-conserving descriptions
\cite{castin_dum_pra_1998, gardiner_pra_1997, girardeau_arnowitt_pr_1959,
girardeau_pra_1998, morgan_jpb_2000}, in which the system is partitioned into
manifestly orthogonal condensate and non-condensate parts using the
Penrose-Onsager criterion \cite{penrose_onsager_pr_1956}, are, in principle,
ideally suited to this regime. However, until recently only the first-order
number-conserving description of Gardiner \cite{gardiner_pra_1997} and Castin
and Dum \cite{castin_dum_pra_1998} offered a description which was numerically
tractable in fully time-dependent form.  Application of this first-order
number-conserving description to the $\delta$-kicked-rotor BEC
\cite{zhang_etal_prl_2004, liu_etal_pra_2006, reslen_etal_pra_2008} and
$\delta$-kicked-harmonic-oscillator BEC \cite{gardiner_etal_pra_2000,
rebuzzini_etal_pra_2007} revealed a general tendency for rapid, unbounded
growth. Unfortunately, as the equation describing the condensate in this
first-order treatment is simply the GPE, this rapid, unbounded growth can be
viewed as a consequence of linearized instabilities present in the GPE, which
would be removed by a higher-order description containing a self-consistent
back-action of the non-condensate on the condensate \cite{gardiner_jmo_2002}.
Thus, similar to the methods discussed above, the first-order description is
also of limited use in describing the long-time dynamics of low-temperature,
driven BEC systems.

However, such a self-consistent back-action \textit{is} present within a
second-order number-conserving description, such as presented by Gardiner and
Morgan in Ref.~\cite{gardiner_morgan_pra_2007} and previously used highly
successfully by Morgan \cite{morgan_etal_prl_2003, morgan_pra_2004,
morgan_pra_2005} to calculate, using a linear response treatment, the
excitation frequencies of a BEC at finite temperature as measured in
experiments at JILA \cite{jin_etal_prl_1996, jin_etal_prl_1997} and MIT
\cite{mewes_etal_prl_1996, stamper-kurn_etal_prl_1998}.  Recently
\cite{billam_gardiner_njp_2012}, the first fully time-dependent implementation
of the second-order number-conserving equations of motion was applied to an
initially zero-temperature $\delta$-kicked-rotor BEC. This application revealed
that the second-order description's self-consistent back-action does indeed
damp out the unbounded growth in non-condensate fraction seen in the
first-order description.  Consequently, the time-dependent form of the
second-order number-conserving description provides an excellent tool for
studying the non-equilibrium dynamics of driven BECs at low temperatures.

The purpose of this paper is primarily to present, in detail, analytic and
numerical techniques which allow one to evolve the second-order
number-conserving equations of motion at finite temperatures, enabling a
systematic exploration of non-equilibrium BEC dynamics at finite temperatures.
A restricted subset of these techniques, limited to zero-temperature initial
conditions, were used to obtain the results in
Ref.~\cite{billam_gardiner_njp_2012}, but were not described in detail in that
work.  We also use these techniques to conduct a finite-temperature
study of the non-equilibrium dynamics of the $\delta$-kicked-rotor BEC,
extending the results of Ref.~\cite{billam_gardiner_njp_2012} and previous
works.  Remarkably, we observe that many of the sharp nonlinear resonance resonance features observed at zero temperature in
Refs.~\cite{billam_gardiner_njp_2012, monteiro_etal_prl_2009} are qualitatively
preserved at temperatures sufficient to cause significant initial condensate
depletion. In particular, our exploration of the effects of initial temperature
predicts a finite-temperature shift in the system's resonance frequencies which
could be experimentally measured and verified.

The remainder of the manuscript is structured as follows:
We begin by introducing, in \secreft{section:nc_description}, the second-order
number-conserving description of the dynamics and present the second-order
equations of motion in their most general form. We give a detailed discussion
of the self-consistent properties and regime of validity of the resulting
description at both zero and finite temperatures. In particular, we discuss
potential problems finding equilibrium initial conditions at higher
temperatures. In \secreft{section:num_imp} we develop a numerical method to
evolve initial states according to the second-order equations of motion.  This
method explicitly includes the nonlinear, non-local terms coupling the
condensate and the quasiparticle modes. It is these terms which maintain
orthogonality between the condensate and non-condensate, and which are
particularly difficult to deal with in the context of general numerical
integration schemes.  In \secreft{section:dkr} we apply the method to
systematically explore the effects of finite initial temperatures in the
$\delta$-kicked-rotor BEC. We also extend the analysis of this system given in
Ref.~\cite{billam_gardiner_njp_2012} by further exploring the contrast between
first- and second-order descriptions, and analyzing the evolution of the
single-particle von Neumann entropy. We predict a shift of resonance
frequencies at finite temperature which could feasibly be detected in
experiments. Section \ref{section:conc} comprises the conclusions. A technical appendix follows, in which we present details of our numerical technique.

\section{Second-order, number-conserving theoretical description\label{section:nc_description}}

\subsection{Motivation\label{section:nc_motivation}}
We consider a system of $N$ bosonic atoms of mass $m$, confined by an external
potential $V(\vx,t)$, and interacting via pair-wise $s$-wave contact
interactions. The Hamiltonian for such a system is given by
\begin{equation}
\label{eqn:bose_gas_hamiltonian_contact_int_recap}
 \hat{H} =  \int d\vx \, \fpsid \left[ H_\mathrm{sp}(\vx,t) + \frac{U_0}{2} \fpsid \fpsi \right] \fpsi \,,
\end{equation}
where $\fpsi$ and $\fpsid$ are second-quantized field operators obeying
standard equal-time bosonic commutation relations. Here, the single-particle
Hamiltonian is
\begin{equation}
H_\mathrm{sp}(\vx,t) = -\frac{\hbar^2}{2m} \nabla^2 + V(\vx,t)\,,
\end{equation}
where
\begin{equation}
U_0 = \frac{4 \pi \hbar^2 a_s}{m}\,,
\end{equation}
for $s$-wave scattering length $a_s$.

Computing the full many-body dynamics of such a system directly from the
Hamiltonian \eqnreft{eqn:bose_gas_hamiltonian_contact_int_recap} is, in
general, an intractable problem. However, in the case where $N$ is large and
the majority of the system is Bose-Einstein condensed, one can obtain an
approximation to the full many-body dynamics via a perturbative description, in
which the non-condensate fraction constitutes the small parameter.  To develop
such a description, we adopt the definition of Bose-Einstein condensation ---
applicable to a finite-size and interacting system as described by
\eqnreft{eqn:bose_gas_hamiltonian_contact_int_recap} --- given by Penrose and
Onsager \cite{penrose_onsager_pr_1956}. In this definition the condensate mode,
$\phi_c(\vx,t)$ is identified as a single macroscopically-occupied eigenstate
of the single-particle density matrix\footnote{Note that in
\eqnreft{eqn:spdm_def} the brackets $\langle \cdots \rangle$ denote an
expectation value with respect to the full many-particle density matrix; at
finite-temperature this involves a thermal, as well as a quantum, average.}
\begin{equation}
\label{eqn:spdm_def}
\rho(\vx, \vy, t) = \left\langle \hat{\Psi}^\dagger(\vy) \hat{\Psi}(\vx) \right\rangle\,.
\end{equation}
Consequently, the condensate mode $\phi_c(\vx,t)$ satisfies
\begin{equation}
\label{eqn:spdm_eigs}
\int d\vy\, \rho(\vx,\vy,t)\phi_c(\vy,t) = N_c(t) \phi_c(\vx,t)\,,
\end{equation}
where the condensate mode occupation, $N_c(t)$, is taken to be much greater
than all other single-particle mode occupations [given by the other eigenvalues
of $\rho(\vx,\vy,t)$] at any time.  

Using this definition, one can explicitly partition the field operator into a
condensate part, $\cannil \phi_c(\vx,t)$, and a non-condensate part, $\dfpsi$:
\begin{equation}
\label{eqn:nc_partition_recap}
 \fpsi = \cannil \phi_c(\vx,t) + \dfpsi \,.
\end{equation}
Here, the condensate annihilation operator $\cannil$ annihilates an atom in the
condensate mode $\phi_c(\vx,t)$. To maintain hermiticity of the single-particle density
matrix one must ensure that the part of the field operator describing the condensate mode, $\cannil \phi_c(\vx,t)$, is explicitly
orthogonal to the part describing the non-condensate, $\dfpsi$, at all times. Formally, one defines 
\begin{equation}
\label{eqn:pedantic_non_condensate_definition}
\dfpsi = \int d\vy\, \mathcal{Q}(\vx,\vy) \hat{\Psi}(\vy,t) \,,
\end{equation}
where
\begin{equation}
\label{eqn:q_def}
\mathcal{Q}(\vx,\vy) = \delta(\vy-\vx) - \phi_c(\vx) \phi_c^\ast(\vy)\,,
\end{equation}
and hence one sees that $\phi_c(\vx,t)$ and $\dfpsi$ are ``orthogonal'' in the sense that
\begin{equation}
\label{eqn:gen_orthog}
\int \, d\vx \phi^*_c(\vx,t) \dfpsi =0\,.
\end{equation}
This partition is
\textit{number-conserving}, in the sense that the system remains in a state of
fixed total atom number.

This \textit{number-conserving} partition can be contrasted the
\textit{symmetry-breaking} partition used in other finite temperature
descriptions, which is based on a partition of the field operator into a finite
expectation value (or order parameter), $\Psi(\vx,t) = \avg{\fpsi}$, and an
operator-valued fluctuation $\hat{\delta}(\vx,t)$,
\begin{equation}
\label{eqn:ssb_partition}
 \fpsi = \Psi(\vx,t) + \hat{\delta}(\vx,t)\,.
\end{equation}
Two related consequences of the symmetry-breaking partition
\eqnrefp{eqn:ssb_partition} are a breaking of the overall $U(1)$ phase-symmetry
of the Hamiltonian, and a general lack of orthogonality between the condensate
and non-condensate parts of the system. Specifically, having constructed a
perturbative description in terms of $\Psi(\vx,t)$ and $\hat{\delta}(\vx,t)$
one generically finds that, in the case of an inhomogeneous order parameter,
$\Psi(\vx,t)$ and $\hat{\delta}(\vx,t)$ are not orthogonal in the sense of
\eqnreft{eqn:gen_orthog}. As a result, one cannot uniquely reassemble a
Hermitian density matrix for the system from $\Psi(\vx,t)$ and
$\hat{\delta}(\vx,t)$, as one can from $\phi_c(\vx,t)$ and $\dfpsi$.  This lack
of an explicit, time-independent orthogonality between condensate and
non-condensate [in the sense of \eqnreft{eqn:gen_orthog}] also means that one
cannot unambiguously determine the condensate population, $N_c$, and the
non-condensate population, $N_t$, associated with the symmetry-breaking
partition. In the context of driven low-temperature BECs, where the dynamics
are extremely strongly influenced by changes in $N_c$, this ambiguity in the
condensate population constitutes a genuine disadvantage of the
symmetry-breaking approach.

A number-conserving description of the system's dynamics is obtained by
treating the non-condensate part $\dfpsi$ perturbatively, by means of an
expansion in terms of a suitable fluctuation operator \cite{gardiner_pra_1997,
castin_dum_pra_1998, gardiner_morgan_pra_2007}.  Neglecting the non-condensate
completely yields a zeroth-order description in the form of the
Gross--Pitaevskii equation (GPE); this takes the form [we explicitly denote the
condensate mode in this zeroth-order description by $\phi_c^{(0)}(\vx,t)$ for
clarity in later sections]
\begin{equation}
\label{eqn:gpe}
i \hbar \frac{\partial \phi_c^{(0)} (\vx,t)}{\partial t} = \left[H_\mathrm{sp}(\vx,t) + \tilde{U} |\phi_c^{(0)}(\vx,t)|^2 -\lambda_0^{(0)} (t) \right] \phi_c^{(0)}(\vx,t)\,,
\end{equation}
where $\tilde{U} = U_0 N_c$, and $\lambda_0^{(0)} (t) $ is given by
\begin{equation}
\label{eqn:gpe_eig_pedantic}
\begin{split}
\lambda_0^{(0)} (t) = \int & d\vx \phi_c^{(0)^\ast}(\vx,t) \left[H_\mathrm{sp}(\vx,t) \right. \\
&\left.+ \tilde{U} |\phi_c^{(0)}(\vx,t)|^2 - i \hbar \frac{\partial}{\partial t} \right] \phi_c^{(0)}(\vx,t)\,.
\end{split}
\end{equation}
As in Refs.~\cite{billam_gardiner_njp_2012, gardiner_morgan_pra_2007,
gardiner_pra_1997}, $\lambda_0^{(0)} (t)$ is defined in such a way as to be generally
time-dependent. However, as $\lambda_0^{(0)} (t)$ is explicitly real, the dynamics
resulting from its time evolution consist only of an irrelevant global phase.
When considering, as we do in this paper, the evolution of a stationary,
equilibrium initial condition subject to a time-dependent perturbation, it is
most convenient to work with a \textit{time-independent} GPE eigenvalue
$\lambda_0^{(0)}$ given by
\begin{equation}
\label{eqn:gpe_eig}
\lambda_0^{(0)} = \int d\vx \phi_c^{(0)^\ast}(\vx,0) \left[H_\mathrm{sp}(\vx,0) + \tilde{U} |\phi_c^{(0)}(\vx,0)|^2 \right] \phi_c^{(0)}(\vx,0)\,,
\end{equation}
where $\phi_c^{(0)}(\vx,0)$ represents the $t=0$ stationary, equilibrium state
of the GPE. Since several subtly different eigenvalues $\lambda$ appear in the
subsequent development, here we have introduced the convention that the
\textit{subscript} index to $\lambda$ denotes the order (with respect to the
non-condensate fluctuation operators) of the \textit{functional} defining
$\lambda$, while the bracketed \textit{superscript} index to $\lambda$ denotes
the order of approximation of the condensate \textit{wavefunction} appearing
inside the functional definition. Furthermore, if any eigenvalue is time
dependent we \textit{always} denote this explicitly with a $(t)$ argument. In
the case where a time argument is absent, the corresponding eigenvalue should
be understood as having been evaluated for the appropriate equilibrium initial
condition.

To work with the time-independent eigenvalue $\lambda_0^{(0)}$ in the GPE
description, one simply makes the replacement $\lambda_0^{(0)} (t) \rightarrow
\lambda_0^{(0)}$ in \eqnreft{eqn:gpe}.  Note that while $\lambda_0^{(0)}$
arises naturally as a nonlinear eigenvalue during the development of a
zeroth-order number-conserving equation of motion, at this level of
approximation it is equivalent to the chemical potential, $\mu$, which would be
introduced as a Lagrange multiplier to determine the average particle number in
a symmetry-breaking approach.

In contrast to the GPE, the second-order number-conserving description we use
in this paper provides a description of both the condensate and the
non-condensate, and consists of mutually-coupled equations for both, which we
outline in the following section.

\subsection{Equations of motion\label{section:eoms}}

Conducting a self-consistent, second-order, number-conserving expansion --- as
detailed in Ref.~\cite{gardiner_morgan_pra_2007}--- leads to a
number-conserving generalized GPE (GGPE) for the dynamics of the condensate
mode $\phi_c(\vx,t)$:
\begin{multline}
 \label{eqn:ggpe_first}
 i \hbar \frac{\partial \phi_c(\vx)}{\partial t} = \left[ H_\mathrm{sp}(\vx)   - \lambda_2^{(2)}(t)  \right] \phi_c(\vx) \\ 
+ \tilde{U} \left[ \left(1-\frac{1}{N_c}\right) |\phi_c(\vx)|^2 + 2 \frac{\tilde{n}(\vx,\vx)}{N_c} \right] \phi_c(\vx)
+ \tilde{U} \phi_c^\ast(\vx)  \frac{\tilde{m}(\vx,\vx)}{N_c} \\
- \tilde{U} \int d\vy\, |\phi_c(\vy)|^2 \left( \frac{\tilde{n}(\vx,\vy)}{N_c} \phi_c(\vy) 
+ \phi_c^\ast(\vy) \frac{\tilde{m}(\vx,\vy)}{N_c}\right)\,.
\end{multline}
Here we have introduced the convention, adopted generally hereafter, of
omitting explicit time-dependences wherever this aids clarity. The dynamics of the
non-condensate enter the GGPE \eqnrefp{eqn:ggpe_first} through the normal
[$\tilde{n}(\vx,\vy)$] and anomalous [$\tilde{m}(\vx,\vy)$] averages. These are
respectively defined by
\begin{equation}
\tilde{n}(\vx,\vy) = \avg{\tilde{\Lambda}^\dagger(\vy) \tilde{\Lambda}(\vx)}\,,
\label{normal}
\end{equation}
and 
\begin{equation}
\tilde{m}(\vx,\vy) = \avg{\tilde{\Lambda}(\vy) \tilde{\Lambda}(\vx)}\,,
\label{anomalous}
\end{equation}
where 
\begin{equation}
 \tilde{\Lambda} (\vx) = \frac{1}{\sqrt{N_c}} \hat{a}_c^\dagger \delta \hat{\Psi}(\vx) \,,
\label{eqn:nc_def}
\end{equation}
is the number-conserving fluctuation operator in which a perturbative expansion
has been conducted. Note that the corresponding small parameter is
$\sqrt{N_t(t)/N_c(t)}$, where $N_t(t) = N - N_c(t)$ is the occupation of the
non-condensate.  We have also introduced the \textit{complex, time-dependent}
``GGPE eigenvalue'' $\lambda_2^{(2)}(t)$, which is given by
\begin{multline}
\label{eqn:ggpe_eig_first}
\lambda_2^{(2)}(t) = \tilde{U} \int d\vx\, \phi_c^\ast(\vx)^2 \frac{\tilde{m}(\vx,\vx)}{N_c}  \\ 
+ \int d\vx\, \phi_c^\ast(\vx) \left\{H_\mathrm{sp}(\vx) + \tilde{U} \left[\left(1-\frac{1}{N_c} \right)|\phi_c(\vx)|^2 \right. \right. \\ 
\left. \left. +2 \frac{\tilde{n}(\vx,\vx)}{N_c} \right] -i\hbar  \frac{\partial}{\partial t}  \right\} \phi_c(\vx)  \,.
\end{multline}
It is essential to emphasize that this ``eigenvalue'' is in general
time-dependent, and that \eqnreft{eqn:ggpe_eig_first} should be taken to be
evaluated using the values of $\phi_c(\vx)$, $N_c$, etc. at a particular
time. As we demonstrate in the subsequent section, this time dependence of
$\lambda_2^{(2)} (t)$ is crucial in order to obtain self-consistent number dynamics.
However, as in the case of the GPE \eqnsrefp{eqn:gpe}{eqn:gpe_eig} it is
considerably more convenient to work with a \textit{real, time-independent}
eigenvalue, and in \secreft{section:num_imp} we will reformulate the
second-order equations of motion in order to consistently (up to second order) replace $\lambda_2^{(2)}(t)$ with
such a quantity.

The GGPE must be coupled to a consistent set of equations of motion for the
non-condensate; at second-order these consist of the number-conserving modified
Bogoliubov--de Gennes equations (MBdGE):
\begin{multline}
\label{eqn:bdg_first}
i \hbar \frac{\partial}{\partial t} \bdgvector{\vx} =\\ \int d\vy\, \left( \begin{array}{cc} \mathcal{L}(\vx,\vy) &  \mathcal{M}(\vx,\vy) \\ -\mathcal{M}^\ast (\vx,\vy) &  -\mathcal{L}^\ast (\vx,\vy)  \end{array} \right) \bdgvector{\vy}\,,
\end{multline}
where
\begin{multline}
 \mathcal{L}(\vx ,\vy)  = \delta(\vx-\vy) \left[ H_\mathrm{sp}(\vy) + \tilde{U} \vert \phi_c(\vy) \vert^2 - \lambda_0^{(2)} (t) \right] \\
  + \int d\vz\, \mathcal{Q}(\vx,\vz) \tilde{U} \vert \phi_c(\vz)\vert^2 \mathcal{Q}(\vz,\vy) \,,
\end{multline}
and
\begin{equation}
 \mathcal{M}(\vx,\vy) = \int d\vz\, \mathcal{Q}(\vx,\vz) \tilde{U} \phi(\vz)^2 \mathcal{Q}^\ast(\vz,\vy) \,.
\end{equation}
The modification in these equations --- with respect to the ordinary BdG equations
obtained in a symmetry-breaking description --- is the appearance of the projector
$\mathcal{Q}(\vx,\vy)$ which explicitly enforces the orthogonality of the condensate and
non-condensate, and of the time-dependent ``GPE eigenvalue'' $\lambda_0^{(2)} (t)$. This ``GPE
eigenvalue'' is obtained by substituting the GGPE condensate mode $\phi_c(\vx,t)$
into the GPE \eqnrefp{eqn:gpe} in place
of the GPE condensate mode $\phi_c^{(0)}(\vx,t)$, and allowing the eigenvalue to be generically time-dependent. That is,
\begin{equation}
\label{eqn:gpe_eig_ggpe_pedantic}
\lambda_0^{(2)} (t) = \int d\vx \phi_c^{\ast}(\vx,t) \left[H_\mathrm{sp}(\vx,t) + \tilde{U} |\phi_c(\vx,t)|^2 - i \hbar \frac{\partial}{\partial t} \right] \phi_c(\vx,t)\,,
\end{equation}
where we have resurrected time-arguments for clarity.
As in the zeroth-order, GPE, case ---  where it was possible, and convenient
for the case of a stationary, equilibrium initial state, to replace the
time-dependent GPE eigenvalue $\lambda_0^{(0)} (t)$ with a time-independent GPE
eigenvalue $\lambda_0^{(0)}$ --- $\bar{\lambda}_0 (t)$ is an explicitly real
quantity. It will thus later be convenient, for the case of a stationary,
equilibrium initial state, to replace the time-dependent ``GPE eigenvalue''
$\lambda_0^{(2)} (t)$ with the time-independent ``GPE eigenvalue''
$\lambda_0^{(2)}$ given by
\begin{equation}
\label{eqn:gpe_eig_ggpe_pedantic_time_independent}
\lambda_0^{(2)} = \int d\vx \phi_c^{\ast}(\vx,0) \left[H_\mathrm{sp}(\vx,0) + \tilde{U} |\phi_c(\vx,0)|^2 \right] \phi_c(\vx,0)\,,
\end{equation}
Note, however, that the replacement of $\lambda_0^{(2)} (t)$ with
$\lambda_0^{(2)}$ must be accomplished consistently with the replacement
of $\lambda_2^{(2)} (t)$ with a real, time-independent quantity as discussed previously. The details of this consistent replacement are outlined in
\secreft{section:num_imp}.

The generalized Gross--Pitaevskii equation \eqnrefp{eqn:ggpe_first} and the
modified Bogoliubov--de Gennes equations \eqnrefp{eqn:bdg_first} complete the
fully dynamical, second-order, number-conserving description obtained by
Gardiner and Morgan in Ref.~\cite{gardiner_morgan_pra_2007} and previously
used within a linear response treatment by Morgan \cite{morgan_etal_prl_2003,
morgan_pra_2004, morgan_pra_2005}. In this paper, we develop these equations
into a form where their fully dynamical time evolution can be realized numerically,
as in Ref.~\cite{billam_gardiner_njp_2012}.  

\subsection{Discussion\label{section:eom_discuss}}
Before we outline our method for the simultaneous numerical solution of
\eqnsreft{eqn:ggpe_first}{eqn:bdg_first}, a few comments regarding the
properties of the second-order equations of motion and their regime of validity
are in order. Firstly we note that, in general, the diagonal part of the
anomalous average, $\tilde{m}(\vx,\vx)$, is ultraviolet-divergent and must be
appropriately renormalized; this procedure is described in detail in
Refs.~\cite{gardiner_morgan_pra_2007} and \cite{morgan_jpb_2000}. One exception
is in the case of quasi-one-dimensional (quasi-1D) systems, such as the one we consider in
\secreft{section:dkr}, where renormalization is not necessary.

Secondly, the number-conserving equations of motion used here are derived by
expanding the total Hamiltonian in powers of the number-conserving fluctuation
operator $\tilde{\Lambda}(\vx)$ \eqnrefp{eqn:nc_def}.  This operator is
advantageous for three primary reasons: (1) it scales proportionally to the
number of non-condensate atoms, which we wish to treat as a small parameter;
(2) it avoids the need to expand inverse-square-root number-operators when
expanding the Hamiltonian, which is particularly convenient; and (3) it is a
well-defined fluctuation operator in the sense that
$\avg{\tilde{\Lambda}(\vx,t)}=0$.  It should be noted that the commutation
relations of $\tilde{\Lambda}(\vx)$
\begin{equation}
\label{eqn:approx_bosonic}
\begin{split}
 \com{\tilde{\Lambda} (\vx) }{\tilde{\Lambda}^\dagger (\vy)} = &\frac{\hat{N}_c}{N_c} \mathcal{Q}(\vx,\vy) \\
&- \frac{1}{N_c} \delta \hat{\Psi}^\dagger (\vy) \delta \hat{\Psi}(\vx) \,,
\end{split}
\end{equation}
give rise to quasiparticles which are, in principle, only approximately
bosonic. However, when restricted to the (quadratic) order of approximation we
consider, the corresponding quasiparticles are indeed exactly bosonic. Hence,
the favorable properties of $\tilde{\Lambda}(\vx)$ mentioned above make it a
preferable choice for developing the second-order equations of motion
\cite{morgan_jpb_2000, morgan_etal_prl_2003, morgan_pra_2004, morgan_pra_2005,
gardiner_morgan_pra_2007, billam_gardiner_finess_2011}.

Note also that the appearance of higher-than-second-order fluctuation terms in
the equations of motion has been prevented by working within a consistent
Gaussian approximation \cite{gardiner_morgan_pra_2007}; that is, all
\textit{quadratic} products of operators are required to take the form of pair
averages. This constitutes a Gaussian approximation in that all higher-order
moments of the fluctuation distribution are assumed to be describable in terms
of $\avg{\tilde{\Lambda}^\dagger(\vx) \tilde{\Lambda}(\vx)}$ and
$\avg{\tilde{\Lambda}(\vx) \tilde{\Lambda}(\vx)}$
\cite{kohler_burnett_pra_2002}. As was explicitly demonstrated by Morgan
\cite{morgan_jpb_2000} this is consistent within the second-order
number-conserving description (see also Refs.~\cite{gardiner_morgan_pra_2007,
proukakis_jackson_jpb_2008}).

\begin{figure}[t]
\centering
 
\includegraphics[width=3.375in]{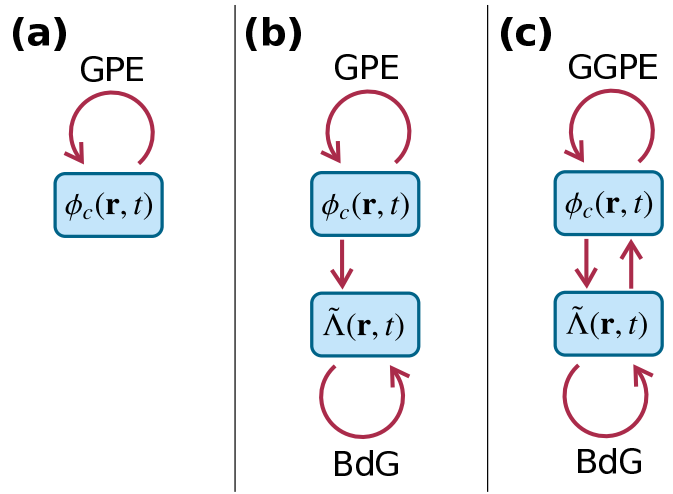}
 \caption{Schematic representation of the zeroth-, first-, and second-order
number-conserving equations of motion. At zeroth order, (a), the non-condensate
is ignored and the condensate mode $\phi_c^{(0)}(\vx)$ is described by the
Gross--Pitaevskii equation (GPE) \eqnrefp{eqn:gpe}.  At first order, (b), the
GPE \eqnrefp{eqn:gpe} is coupled to modified Bogoliubov--de Gennes equations
(MBdGE) \eqnrefp{eqn:bdg_first} for the non-condensate fluctuation operator
$\tilde{\Lambda}(\vx)$. At this order, the evolution of $\tilde{\Lambda}(\vx)$
depends on the evolution of $\phi_c^{(0)}(\vx)$, but the converse is not true;
this order of approximation can be interpreted as treating the condensate as an
infinite atomic reservoir. At second order, (c), the non-condensate is again
described by the MBdGE \eqnrefp{eqn:bdg_first};  however, the evolution of the
condensate mode $\phi_c(\vx)$ is now determined by the generalized
Gross--Pitaevskii equation (GGPE) \eqnrefp{eqn:ggpe_first}. This pairing of
equations produces fully self-consistent number dynamics.\label{fig:eoms}}
\end{figure}

A key property of the second-order number-conserving equations of motion is
their \textit{number-self-consistency}: in contrast to $\lambda_0^{(0)}$ and
$\lambda_0^{(2)}$ --- which can both be considered as low-order
approximations to the chemical potential --- $\lambda_2^{(2)} (t)$ is a
\textit{complex} eigenvalue. The meaning of the imaginary part of
$\lambda_2^{(2)} (t)$ can be understood by considering the (implicit) time
dependence of $N_c$, which is given (to quadratic order) by
\begin{equation}
\begin{split}
i \hbar \frac{d N_c}{d t} &= i \hbar \frac{d}{d t} \left[N-\int d\vx\, \avg{\tilde{\Lambda}^\dagger(\vx) \tilde{\Lambda}(\vx)} \right]\,, \\
&= -\int d\vx\,\left\{ \avg{\tilde{\Lambda}^\dagger(\vx) \left[i\hbar \frac{d}{dt} \tilde{\Lambda}(\vx)\right]} \right.\\
&\phantom{=}\quad +\left. \avg{\left[ i \hbar \frac{d}{dt} \tilde{\Lambda}^\dagger(\vx)\right] \tilde{\Lambda}(\vx)} \right\}\,,\\
&= \tilde{U} \int d\vx\, \left[\phi_c^\ast(\vx)^2 \tilde{m}(\vx,\vx) - \tilde{m}^\ast(\vx,\vx) \phi_c(\vx)^2 \right]\,,\\
&= [\lambda_2^{(2)} (t) - {\lambda_2^{(2)}}^* (t)] N_c \label{eqn:numdyn}\,.
\end{split}
\end{equation}
Thus the time-dependent, imaginary part of $\lambda_2^{(2)} (t)$ acts to keep the
condensate mode $\phi_c(\vx)$ normalized to unity despite the growth or decay
of the condensate population. This illustrates the presence of
number-self-consistency in the dynamical coupling between the GGPE and MBdGE.
In contrast, the first-order description \cite{gardiner_pra_1997,
castin_dum_pra_1998} --- which consists of the same MBdGE
\eqnrefp{eqn:bdg_first} coupled to the ordinary GPE \eqnrefp{eqn:gpe} --- is
\textit{not} in general number-self-consistent, since the condensate population
is fixed. The first-order description can only be considered to be
number-self-consistent when viewed as the limit of the second-order description
as $N \rightarrow \infty$ \cite{gardiner_morgan_pra_2007}. The zeroth-order
description, consisting of the GPE \eqnrefp{eqn:gpe} alone, is trivially
number-self-consistent, as it ignores the growth and decay of the condensate
altogether. The number-self-consistency of the zero-, first-, and second-order
number-conserving descriptions are illustrated schematically in
\figreft{fig:eoms}{}.

A final feature of the second-order, number-conserving equations of motion is
that the non-condensate is described by a MBdGE \eqnrefp{eqn:bdg_first} which does
not contain pair averages of the non-condensate operators; the terms
$\mathcal{M}(\vx,\vy)$ are in no way altered from the first-order description
and the terms $\mathcal{L}(\vx,\vy)$ appearing in the MBdGE consist only of a
GPE Hamiltonian and the ``GPE eigenvalue'' of \eqnreft{eqn:gpe_eig_ggpe_pedantic_time_independent}.
Importantly, however, the GPE Hamiltonian and the ``GPE eigenvalue'' are, in
the second-order description, evaluated in terms of the second-order GGPE
wavefunction $\phi_c(\vx)$.  This leads to an apparent inconsistency in that
the spinors $[\phi_c(\vx),0]^T$ and $[0,\phi_c^\ast(\vx)]^T$ are no longer
exact, zero-energy solutions of the MBdGE, as they would be for the GPE
wavefunction $\phi_c^{(0)}$. This has the unfortunate side-effect of making the
identification of self-consistent initial conditions difficult at high
temperatures, particularly in inhomogeneous systems.  However, this problem can
be viewed purely as a consequence of applying the theory outside of its regime
of validity, as argued by Morgan in his study of condensate excitations at
finite-temperature \cite{morgan_etal_prl_2003, morgan_pra_2004,
morgan_pra_2005}; in this work he demonstrated that the second-order
description remains self-consistent at high temperatures (approaching $N_t \sim
N_c$) provided one restricts oneself to a linear response treatment, in which a
self-consistent equilibrium solution is not necessary. A fully dynamical
treatment, which requires a self-consistent equilibrium initial condition, is
thus restricted to lower temperatures (such that $N_t < N_c$) than a linear
response treatment.

\section{Fully time-dependent numerical implementation\label{section:num_imp}}

\subsection{Reformulation of equations of motion\label{section:eom_reform}}

\subsubsection{Elimination of complex, time-dependent ``eigenvalue''\label{section:eliminate_complex_eig}}

In this Section we develop a numerical method for evolving the combined GGPE
and MBdGE system of \eqnsreft{eqn:ggpe_first}{eqn:bdg_first}.  In order to do
so, it is of great convenience to eliminate the imaginary part of the GGPE
``eigenvalue'' $\lambda_2^{(2)} (t)$ \eqnrefp{eqn:ggpe_eig_first}; this can be done by
describing the condensate using a mode function normalized to the condensate
population, $N_c$. Hence, we define
\begin{equation}
\psi(\vx) = \sqrt{N_c} \phi_c(\vx)\,,
\end{equation}
in terms of which the GGPE can be re-expressed [using \eqnreft{eqn:numdyn} for the
number evolution] as
\begin{multline}
\label{ggpe_generic}
 i \hbar  \frac{\partial \psi(\vx)}{\partial t} = \left[ H_{\rm sp}(\vx) + U_0 |\psi(\vx)|^2 - \bar{\lambda}_2^{(2)} \right] \psi(\vx) \\
+ U_0 \left[\tilde{n}(\vx,\vx) - \frac{|\psi(\vx)|^2}{N_c} \right] \psi(\vx) \\
 + U_0 \tilde{n}(\vx,\vx) \psi(\vx) - \frac{U_0}{N_c} \int d\vy\, \psi(\vy) |\psi(\vy)|^2 \tilde{n}(\vx,\vy) \\
 + U_0 \tilde{m}(\vx,\vx) \psi^\ast (\vx) - \frac{U_0}{N_c} \int d\vy\, \psi^\ast(\vy) |\psi(\vy)|^2 \tilde{m}(\vx,\vy) \,,
\end{multline}
where the adjusted GGPE eigenvalue, $\bar{\lambda}_2^{(2)} = \mathop{\mbox{Re}}(\lambda_2^{(2)})$, is given by
\begin{multline}
\label{eqn:defggpeeig}
\bar{\lambda}_2^{(2)} = \lambda_0^{(2)} + \frac{U_0}{N_c}\int d\vx\, \left\{ \psi^\ast(\vx) \left[ 2\tilde{n}(\vx,\vx) 
- \frac{1}{N_c}|\psi(\vx)|^2 \right] \psi(\vx) \right. \\
\left. + \frac{U_0}{2N_c} \left[{\psi(\vx)^\ast}^2
\tilde{m}(\vx,\vx) + \psi(\vx)^2 \tilde{m}^\ast(\vx,\vx) \right] \right\}\,.
\end{multline}
This is explicitly real, and should be evaluated explicitly in terms of the
stationary, equilibrium initial condition of the GGPE at $t=0$.

Note that in order to cast the GGPE in the above form, with time-independent
eigenvalue $\bar{\lambda}_2^{(2)}$, we have formally moved into a frame which
cancels the time-dependence of the \textit{real} part of $\lambda_2^{(2)} (t)$.
However, transforming the MBdGE into the same frame leads to additional terms
of the form $\delta(\vx-\vy) [f(t=0)-f(t)]$ (where $t=0$ refers to the
equilibrium initial condition) appearing in the operator
$\mathcal{L}(\vx,\vy)$. Specifically, such terms appear with
\begin{multline}
f(t) = \frac{U_0}{N_c}\int d\vx\, \left\{ \psi^\ast(\vx) \left[ 2\tilde{n}(\vx,\vx) 
- \frac{1}{N_c}|\psi(\vx)|^2 \right] \psi(\vx) \right. \\
\left. + \frac{U_0}{2N_c} \left[{\psi(\vx)^\ast}^2
\tilde{m}(\vx,\vx) + \psi(\vx)^2 \tilde{m}^\ast(\vx,\vx) \right] \right\} \,,
\end{multline}
where all quantities except $U_0$ on the right hand side are time-dependent.
Since these terms are small, and formally of the same (cubic) order with
respect to the fluctuation operator $\tilde{\Lambda}(\vx)$ as terms already
omitted from the description of the non-condensate at second order
\cite{gardiner_morgan_pra_2007}, they should be neglected to maintain a
consistent description. We thus cast the MBdGE as
\begin{multline}
i \hbar  \frac{\partial}{\partial t} \tilde{\Lambda}(\vx) = \left[ H_{\rm sp}(\vx) + U_0 |\psi(\vx)|^2 - \lambda_0^{(2)} \right] \tilde{\Lambda}(\vx) \\
 + U_0 |\psi(\vx)|^2 \tilde{\Lambda}(\vx) - \frac{U_0}{N_c} \int d\vy\, \psi^\ast(\vy) |\psi(\vy)|^2 \tilde{\Lambda}(\vy) \psi(\vx) \\
 + U_0 \psi(\vx)^2 \tilde{\Lambda}^\dagger(\vx) - \frac{U_0}{N_c} \int d\vy\, \psi(\vy) |\psi(\vy)|^2 \tilde{\Lambda}^\dagger(\vy) \psi(\vx) \,;
\label{bdg_generic}
\end{multline}
a form which emphasizes the significant structural analogies between the GGPE
and the MBdGE.

\subsubsection{Quasiparticle decomposition\label{section:qp_decomp}}
In numerical studies of non-equilibrium dynamics at finite temperature one
generally wishes to begin from a finite-temperature equilibrium condition and
explore the dynamics resulting from, e.g., driving or an abrupt quench event in
a fully-time dependent way\footnote{One could in principle start from a
non-equilibrium initial condition, but we consider only the equilibrium case
here.}. In the second-order number-conserving description such a thermal and
dynamical equilibrium state corresponds to a self-consistent, stationary
solution of the equations of motion in which the elementary quasiparticle
excitations of the system are populated according to the appropriate thermal
Bose distribution.

The appropriate quasiparticle basis in the second-order number-conserving
description is the basis of Bogoliubov quasiparticle modes which diagonalize
the stationary MBdGE \cite{gardiner_morgan_pra_2007}:
\begin{multline}
\label{eqn:bdg_quasi}
\left( \begin{array}{cc} \mathcal{L}(\vx,\vy) &  \mathcal{M}(\vx,\vy) \\ -\mathcal{M}^\ast (\vx,\vy) & -\mathcal{L}^\ast (\vx,\vy)  \end{array} \right) = \sum_{k=1}^\infty \epsilon_k \left(\begin{array}{c} u_k(\vx) \\ v_k(\vx) \end{array} \right) \left( u_k^\ast(\vy), -v_k^\ast(\vy) \right) \\
- \sum_{k=1}^\infty \epsilon_k \left(\begin{array}{c} v_k^\ast(\vx) \\ u_k^\ast(\vx) \end{array} \right) \left(-v_k(\vy), u_k(\vy) \right) \,.
\end{multline}
In terms of the quasiparticle mode functions $u_k(\vx)$ and $v_k(\vx)$, the
fluctuation operator $\tilde{\Lambda}(\vx)$ can be expanded as
\begin{equation}
\ltwpair{\vx} = \sum_{k=1}^\infty \qannil \uvpair{\vx} + \sum_{k=1}^\infty \qcreate \uvstarpair{\vx}\,,
\label{eqn:quas_expan}
\end{equation}
where $\qcreate$ and $\qannil$ are quasiparticle creation and annihilation
operators which, at this order, are assumed to have bosonic commutation
relations (see \secreft{section:eom_discuss}).  Assuming all time-dependence to
reside in the quasiparticle mode functions $u_k(\vx)$ and $v_k(\vx)$, with the
quasiparticle creation and annihilation operators $\qcreate$ and $\qannil$
time-independent, the MBdGE \eqnrefp{bdg_generic} take the form
\begin{multline}
i \hbar \frac{\partial}{\partial t} u_k(\vx)  = \left[ H_{\rm sp}(\vx) + U_0 |\psi(\vx)|^2 - \lambda_0^{(2)} \right] u_k(\vx) \\
 + U_0 |\psi(\vx)|^2 u_k(\vx) - \frac{U_0}{N_c} \int d\vy\, \psi^\ast(\vy) |\psi(\vy)|^2 u_k(\vy) \psi(\vx) \\
 + U_0 \psi(\vx)^2 v_k(\vx) - \frac{U_0}{N_c} \int d\vy\, \psi(\vy) |\psi(\vy)|^2 v_k(\vy) \psi(\vx) \,,
\label{eqn:bdg_second_1}
\end{multline}
and
\begin{multline}
i \hbar  \frac{\partial}{\partial t} v_k^\ast(\vx) = \left[ H_{\rm sp}(\vx) + U_0 |\psi(\vx)|^2 - \lambda_0^{(2)} \right] v_k^\ast(\vx) \\
 + U_0 |\psi(\vx)|^2 v_k^\ast(\vx) - \frac{U_0}{N_c} \int d\vy\, \psi^\ast(\vy) |\psi(\vy)|^2 v_k^\ast(\vy) \psi(\vx) \\
 + U_0 \psi(\vx)^2 u_k^\ast(\vx) - \frac{U_0}{N_c} \int d\vy\, \psi(\vy) |\psi(\vy)|^2 u_k^\ast(\vy) \psi(\vx) \,.
\label{eqn:bdg_second_2}
\end{multline}

At initial thermal equilibrium, the quasiparticle creation and
annihilation operators have the following pair averages:
\begin{gather}
\label{eqn:qp_pop2}
\avg{\tilde{b}_k^\dagger \tilde{b}_l} = \delta_{kl} N_k\,,\\
\avg{\tilde{b}_k \tilde{b}_l} = \avg{\tilde{b}^\dagger_k \tilde{b}^\dagger_l} = 0,
\end{gather}
where $N_k$ is the Bose-Einstein factor \cite{gardiner_morgan_pra_2007,
morgan_pra_2004, morgan_jpb_2000}
\begin{equation}
\label{eqn:pops}
N_k = \left[ \exp \left(\frac{\epsilon_k - [\mu - \lambda_0^{(2)} ]}{k_B T} \right) - 1 \right]^{-1}\,.
\end{equation}
The term $\mu-\lambda_0^{(2)}$ represents a finite-size correction, given by
\cite{morgan_jpb_2000, morgan_pra_2004, mandl}
\begin{equation}
\mu - \lambda_0^{(2)} = -k_B T \ln \left( 1 - N_c^{-1} \right)\,.
\end{equation}
This leads to quasiparticle expressions for the non-condensate normal and
anomalous averages \eqnsrefp{normal}{anomalous}
\begin{align}
\tilde{n}(\vx,\vy) &= \sum_{k=1}^\infty N_k u_k(\vx)u_k^\ast(\vy) + \sum_{k=1}^\infty (N_k + 1) v_k^\ast(\vx) v_k(\vy) \,, \label{eqn:norm_exp2} \\
\tilde{m}(\vx,\vy) &= \sum_{k=1}^\infty N_k u_k(\vx)v_k^\ast(\vy) + \sum_{k=1}^\infty (N_k + 1) v_k^\ast(\vx) u_k(\vy) \, \label{eqn:anom_exp2}.
\end{align}
We reiterate that, by choosing a scheme where all time dependence resides in
the quasiparticle mode functions $u_k(\vx)$ and $v_k(\vx)$, the quasiparticle
populations $N_k$ appearing in \eqnsreft{eqn:norm_exp2}{eqn:anom_exp2} remain
\textit{fixed} in time. 

Using the above expressions we proceed, in the next Section, to re-cast the
GGPE in terms of the quasiparticle mode functions, and re-cast the combined
GGPE and MBdGE --- now in the form of Eqs.~(\ref{ggpe_generic}),
(\ref{eqn:bdg_second_1}), and (\ref{eqn:bdg_second_2}) --- in the final form which
we will use to conduct a simultaneous numerical solution.

\subsubsection{Introduction of the spinor $\zeta$\label{section:spinor}}

The primary difficultly in simulating the coupled GGPE-MBdGE system numerically
is the problem of orthogonalization: \textit{both} equations contain terms
which function to maintain orthogonality between the condensate and
non-condensate.  This is in contrast to the case of the first-order GPE-MBdGE
system, where the GPE evolves in isolation from the MBdGE; this de-coupling in
the first-order system means the evolution of the MBdGE can be computed by
\textit{ignoring} the projector terms in the MBdGE throughout the evolution,
and simply projecting the final state orthogonally to the condensate
\cite{gardiner_etal_pra_2000}. However, if one were to
similarly ignore the projectors during the evolution of the second-order
GGPE-MBdGE system, one would then have to re-orthogonalize both the condensate
and quasiparticle modes with respect to an \textit{unknown} basis at the end of
the evolution. Consequently, the projection terms, and the non-local integrals
they involve, must be explicitly included in any numerical method. In this
Section we develop such a method, by re-casting the GGPE and MBdGE in a form
which exploits their apparent symmetries, and allows us to include the
projection terms using a split-step technique.

After substituting the quasiparticle expressions for the normal and anomalous
average \eqnsrefp{eqn:norm_exp2}{eqn:anom_exp2} into the GGPE
\eqnrefp{ggpe_generic}, the GGPE can be recast in the form
\begin{multline}
i\hbar \frac{\partial \psi(\vx)}{\partial t} = \left[ H_{\rm GP}(\vx) + B(\vx) \right] \psi(\vx) \\ 
+ \sum_{k=1}^\infty (N_k+1) A_k(\vx) v_k^\ast(\vx) + \sum_{k=1}^\infty N_k A_k^\ast(\vx) u_k(\vx)\,,
\end{multline}
and the MBdGE \eqnsrefp{eqn:bdg_second_1}{eqn:bdg_second_2} can be recast in
the form
\begin{align}
i\hbar \frac{\partial u_k(\vx)}{\partial t} &= H_{\rm GP}(\vx) u_k(\vx) + A_k(\vx) \psi(\vx) \,,\\
i\hbar \frac{\partial v_k^\ast(\vx)}{\partial t} &= H_{\rm GP}(\vx) v_k^\ast(\vx) + A_k^\ast(\vx) \psi(\vx) \,,
\end{align}
where
\begin{gather}
H_\mathrm{GP}(\vx) = H_\mathrm{sp}(\vx) + U_0 |\psi(\vx)|^2 - \lambda_0^{(2)}\,,\\
A_k(\vx) = U_0 \left[ v_k(\vx)\psi(\vx) + u_k(\vx)\psi^\ast(\vx) - I_k \right]\,,\\
I_k = \frac{1}{N_c} \int d\vx\, \left[ v_k(\vx)\psi(\vx) + u_k(\vx)\psi^\ast(\vx) \right] |\psi(\vx)|^2 \,,
\end{gather}
and
\begin{multline}
B(\vx) = U_0 \left[ \sum_{k=1}^\infty N_k|u_k(\vx)|^2 + (N_k+1)|v_k(\vx)|^2 \right] \\
- U_0 \frac{|\psi(\vx)|^2}{N_c} - \lambda^\prime\,,
\end{multline}
where $\lambda^\prime = \bar{\lambda}_2^{(2)} - \lambda_0^{(2)}$.
This reformulation of the problem allows one to write the coupled evolution of
the condensate wavefunction and the first $M$ quasiparticle modes as a
nonlinear matrix equation in a ($2M+1$)-dimensional spinor space:
\begin{equation}
i \hbar \frac{\partial}{\partial t} \zeta(\mathbf{r}) = \Gamma(\vx) \zeta(\mathbf{r})\,.
\end{equation}
Here the vector $\zeta(\mathbf{r})$ is defined by
\begin{equation}
\zeta(\mathbf{r}) = \left[ \psi(\vx), v_1^\ast(\vx), \ldots , v_M^\ast(\vx) , u_1(\vx), \ldots , u_M(\vx) \right]^T\,,
\end{equation}
and the operator $\Gamma(\vx)$ is defined by
\begin{widetext}
\begin{equation}
\Gamma(\vx) =
\left( \begin{array}{cccccccccc}
H_{\rm GP}(\vx) + B(\vx) & (N_1 +1) A_1(\vx) & (N_2 +1) A_2(\vx) &\cdots & (N_M +1) A_M(\vx) & N_1 A_1^\ast(\vx) & N_2 A_2^\ast(\vx) & \cdots & N_M A_M^\ast(\vx) \\
A_1^\ast(\vx) & H_{\rm GP}(\vx) & 0 & \cdots &0 & 0 & 0 & \cdots & 0 \\
A_2^\ast(\vx) & 0 & H_{\rm GP}(\vx) & \cdots &0 & 0 & 0 & \cdots & 0 \\
\vdots & \vdots &\vdots & \ddots & \vdots & \vdots & \vdots & \cdots & \vdots \\
A_M^\ast(\vx) & 0 & 0 & \cdots & H_{\rm GP}(\vx) & 0 & 0 & \cdots & 0 \\
A_1(\vx) & 0 & 0 & \cdots &0 & H_{\rm GP}(\vx) & 0 & \cdots & 0 \\
A_2(\vx) & 0 & 0 & \cdots &0 & 0 & H_{\rm GP}(\vx) & \cdots & 0 \\
\vdots & \vdots & \vdots & \vdots & \vdots & \vdots & \vdots & \ddots & \vdots \\
A_M(\vx) & 0 & 0 & \cdots &0 & 0 & 0 & \cdots & H_{\rm GP}(\vx) \\
\end{array}\right)\,.
\end{equation}
\end{widetext}
In any actual calculation, this spinor space is rendered finite-dimensional by
the need for a finite quasiparticle momentum cut-off $M$. However, $M$ may, in
principle be arbitrarily large.

\subsection{Operator-splitting scheme for time-evolution\label{section:time_splitting}}

\subsubsection{Separation of position and momentum terms\label{section:position_momentum_split}}

As we have already accounted for all creation and annihilation operators
through the quasiparticle decomposition, each entry in the matrix defining
$\Gamma(\vx)$ can be thought of as an operator in the first-quantized
sense\footnote{That is, an operator which could appear on the right side of a
\textit{single-particle} Schr\"{o}dinger equation.}. From an analytic
perspective, this notation seems to achieve little more than `tidying' ---
abstracting away much of the detail. Importantly, however, all the operators
which are off-diagonal in the spinor space [the operators $A_k(\vx)$] are
diagonal in the position representation, since they ultimately consist of a
multiplication by a spatially-varying function. In contrast, all the operators
which have off-diagonal components in the position representation [that is, the
kinetic energy operator implicitly contained in $H_\mathrm{sp}(\vx)$ and hence
in $H_\mathrm{GP}(\vx)$] appear only on the diagonal in the spinor space.

From a numerical perspective this arrangement is extremely useful, as it makes the
evolution amenable to a split-step approximation.  This is achieved by
splitting $\Gamma(\vx)$ into the sum of a term representing the linear
single-particle evolution, $\Gamma_\mathrm{L}(\vx)$, and a term representing
nonlinear parts of the evolution, $\Gamma_\mathrm{N}(\vx)$. The linear,
single-particle term is defined by
\begin{equation}
\Gamma_\mathrm{L}(\vx) = I \otimes H_\mathrm{L}(\vx)\,,\\
\end{equation}
where $I$ represents the $(2M+1)\times(2M+1)$ identity matrix,
and we have also defined
\begin{equation}
H_\mathrm{L}(\vx) = H_\mathrm{sp}(\vx) -\lambda_0^{(2)} \,.
\end{equation}
The nonlinear term is then defined simply by
\begin{equation}
\Gamma_\mathrm{N}(\vx) = \Gamma(\vx) - \Gamma_\mathrm{L}(\vx)\,.
\end{equation}
Note that this matrix contains diagonal entries of the form $H_\mathrm{N}(\vx)$
and $H_\mathrm{N}(\vx) + B(\vx)$ where
\begin{equation}
H_\mathrm{N}(\vx) = H_\mathrm{GP}(\vx) - H_\mathrm{L}(\vx)\,.
\end{equation}
Written in this form, $\Gamma_\mathrm{L}(\vx)$ contains all kinetic energy
terms; while these terms are not diagonal in the position representation, they
do all lie on the diagonal in the spinor space. Consequently, the evolution due
to the kinetic energy terms can be computed for each mode function
\textit{separately}. In contrast, while $\Gamma_\mathrm{N}(\vx)$ does contain
off-diagonal elements in the spinor space, each element of
$\Gamma_\mathrm{N}(\vx)$ is diagonal in the position representation.
Consequently, the evolution due to these terms can be computed
straightforwardly over a discrete spatial grid.

In the spinor space, the split-step approximation for the evolution of the
system over short times $\delta t$ is given by 
\begin{multline}
\zeta(\mathbf{r},t+\delta t) = e^{-i\Gamma(\vx) \delta t / \hbar} \zeta(\mathbf{r},t) \\
\approx e^{-i \Gamma_\mathrm{N}(\vx) \delta t / 2\hbar} e^{-i \Gamma_\mathrm{L}(\vx) \delta t / \hbar} e^{-i \Gamma_\mathrm{N}(\vx) \delta t / \hbar} \zeta(\mathbf{r},t)\,,
\label{eqn:split_step_def}
\end{multline}
where each of the three evolution operators on the right is assumed to act
instantaneously, and the symmetrization reduces the error to the order of
$\delta t^3$ \cite{javanainen_ruostekoski_jpa_2006}. Since $\Gamma_\mathrm{L}(\vx)$ is diagonal in the
spinor space, we have
\begin{equation}
e^{-i\Gamma_\mathrm{L}(\vx)\delta t / \hbar} = I \otimes e^{-iH_\mathrm{L}(\vx)\delta t / \hbar}\,.
\end{equation}

\subsubsection{Analytic form for position-space evolution\label{section:split_step}}
The spatial operator $\Gamma_\mathrm{N}(\vx)$, is more problematic because it
is not diagonal in the spinor space, and because it contains the nonlocal,
nonlinear terms necessary to preserve orthogonality between the condensate and
non-condensate.  However, each of its entries \textit{is} diagonal in the
position representation, meaning we only need exponentiate
$\Gamma_\mathrm{N}(\vx)$ in the spinor space in order to obtain an operator we
can evaluate and use for short-time propagation \textit{at a given position}:
such an operator --- in distinct contrast to $e^{-i\Gamma_\mathrm{L}(\vx)\delta
t / \hbar}$ which consists of an application of $e^{-iH_\mathrm{L}(\vx)\delta t
/ \hbar}$ to each mode \textit{individually} --- \textit{couples} the mode
functions by acting on \textit{all modes at once}.  The numerical advantage of
casting the evolution in terms of such an operator is that it almost entirely
overcomes the problem of retaining orthogonality between the condensate and the
quasiparticle modes, which poses severe difficulties for other schemes.  In
practice we do find that a correction for numerical round-off error is still
required in order to preserve complete orthogonality over long times; this can
be achieved by explicitly orthogonalizing the quasiparticle modes with respect
to the condensate at the end of each time-step.  However, our experience is
that applying such a correction after every time-step does not cause decay of
the total atom number, indicating that the need for such a correction does
indeed stem from the numerical round-off error inherent in finite-precision
arithmetic. Indeed, the problem of loss of orthogonality between vectors due to
finite-precision arithmetic is well-known in the context of, e.g., the
Gram-Schmidt process \cite{golub}.

In principle, the evolution due to the term $e^{-i\Gamma_\mathrm{N}(\vx)\delta
t/2\hbar}$ can be obtained, for a given $\vx$, by exponentiating the matrix
$\Gamma_\mathrm{N}(\vx)$ numerically at each spatial grid point. However, with
the necessary number of floating point operations necessary to diagonalize the
matrix $\Gamma_\mathrm{N}(\vx)$ scaling at least as $(2M+1)^3$ \cite{golub},
this leads to an algorithm which is too slow to be practicable.
To overcome this limitation, we have computed a general analytic expression for
$e^{-i\Gamma_\mathrm{N}(\vx)\delta t/2\hbar}$ for arbitrary $M$: this reduces
the scaling of the computational effort to $(2M+1)^2$ (i.e., equivalent to a
matrix-vector multiplication). This expression is explicitly derived in
Appendix \ref{section:app}, where we also show that, at any specific point
$\vx$ (and time $t$) on the computational grid, the time evolution due to
$\Gamma_\mathrm{N}$ can be approximated locally, after numerical evaluation of
the non-local integrals appearing in the functions $A_k(\vx)$ over the entire
position-space grid, by the following $2M+1$ coupled equations:
\begin{multline}
\psi(\vx,t+\delta t/2) = \left(T_{\rm cos}(\vx,t) -\frac{1}{2} T_{\rm sin}(\vx,t) B(\vx,t) \right)\psi(\vx,t) \\
- T_{\rm sin}(\vx,t) \Xi(\vx,t)\,,
\label{eqn:psi_evo}
\end{multline}
\begin{multline}
v^\ast_k(\vx,t+\delta t/2) = -A_k^\ast(\vx,t) T_{\rm sin}(\vx,t) \psi(\vx,t) + T_{\rm exp}(\vx,t) v_k^\ast(\vx,t) \\
+ A_k^\ast(\vx,t) T_{\rm mix}(\vx,t) \Xi(\vx,t) \,,
\label{eqn:v_evo}
\end{multline}
\begin{multline}
u_k(\vx,t+\delta t/2) =  -A_k(\vx,t) T_{\rm sin}(\vx,t) \psi(\vx,t) + T_{\rm exp}(\vx,t) u_k(\vx,t) \\
+ A_k(\vx,t) T_{\rm mix}(\vx,t) \Xi(\vx,t)\,,
\label{eqn:u_evo}
\end{multline}
where
\begin{equation}
\label{eqn:main_xi_def}
\Xi(\vx,t) = \sum_{l=1}^M (N_l+1)A_l(\vx,t)v_l^\ast(\vx,t) + N_l A_l^\ast(\vx,t) u_l(\vx,t)\,,
\end{equation}
and we have defined
\begin{multline}
\shoveright{T_{\rm exp}(\vx,t) = \exp \left(\frac{-iH_\mathrm{N}(\vx,t) \delta t}{2\hbar}\right)\,,}
\end{multline}
\begin{multline}
T_{\rm cos}(\vx,t) = \exp \left(\frac{-i\left[H_\mathrm{N}(\vx,t)+\frac{1}{2}B(\vx,t)\right]\delta t}{2\hbar}\right) \\
\times \cos \left(\left[\left(\frac{B(\vx,t)}{2}\right)^2 + \Sigma(\vx,t)\right]^{\frac{1}{2}} \frac{\delta t}{2\hbar} \right) \,,
\end{multline}
\begin{multline}
T_{\rm sin}(\vx,t) = \exp \left(\frac{-i\left[H_\mathrm{N}(\vx,t)+\frac{1}{2}B(\vx,t)\right] \delta t}{2\hbar}\right) \\ 
\times \frac{i\sin \left(\left[\left(\frac{B(\vx,t)}{2}\right)^2 + \Sigma(\vx,t)\right]^{\frac{1}{2}} \frac{\delta t}{2\hbar} \right)}
{\left[\left(\frac{B(\vx,t)}{2}\right)^2 + \Sigma(\vx,t)\right]^{\frac{1}{2}}} \,,
\end{multline}
\begin{multline}
T_{\rm mix}(\vx,t) = \frac{T_{\rm cos}(\vx,t) + \frac{1}{2}T_{\rm sin}(\vx,t) B(\vx,t) - T_{\rm exp}(\vx,t)}{\Sigma(\vx,t)}\,,
\end{multline}
and
\begin{equation}
\Sigma(\vx,t) = \sum_{k=1}^M (2N_k +1) |A_k(\vx,t)|^2\,.
\end{equation} 
Note that the quasiparticle index $k$ in \eqnsreft{eqn:v_evo}{eqn:u_evo} runs
from $1$ to $M$, and we have chosen to show time-dependences explicitly in
Eqs.~(\ref{eqn:psi_evo}--\ref{eqn:u_evo}), to make the time-dependent nature of
the expression clear. The analytic evolution scheme given by
Eqs.~(\ref{eqn:psi_evo}--\ref{eqn:u_evo}), incorporated into the symmetrized
split-step method of \eqnreft{eqn:split_step_def}, provides a way to
numerically implement the second-order number-conserving description of
\secreft{section:nc_description}. 

\section{Non-equilibrium dynamics of the $\delta$-kicked-rotor BEC at finite temperatures\label{section:dkr}}

\subsection{Overview\label{section:dkr_overview}}

\begin{figure}[t]
\centering
\includegraphics[width=\columnwidth]{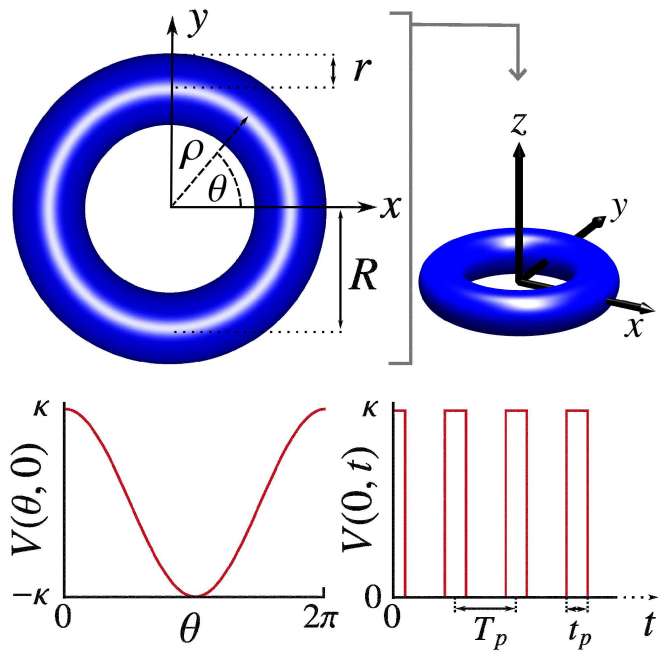}
\caption[The $\delta$-kicked-rotor BEC.]{(color online). The $\delta$-kicked-rotor BEC system:
A repulsively-interacting atomic BEC consisting of $N$ atoms is held in a
quasi-1D, toroidally-shaped trap, and driven by periodic kicks of short
duration $t_p$, which can be modeled as $\delta$-impulses, from a sinusoidal
driving potential. The free parameters of the system are the dimensionless
kick strength $\kappa$, kicking period $T_p$, and interaction strength $g_T$.
See \secreft{section:dkr_system} for a full definition of these parameters.}
\label{fig_system}
\end{figure}

In this Section we apply the split-step numerical method developed in the
previous Section for evolving the second-order, number-conserving equations of
motion --- consisting of coupled GGPE and MBdGE --- to the
$\delta$-kicked-rotor BEC. This system consists of a quasi-1D,
toroidally-trapped, repulsively-interacting atomic BEC driven by $\delta$-kicks
from a spatial cosine potential, and serves as a prototypical example of a
system where condensate depletion occurs dynamically as a result of external
driving. The free parameters of this system can be expressed in terms of
dimensionless kick strength $\kappa$, kick period $T_p$, and interaction
strength $g_T$, plus the total atom number $N$, as shown in
\figreft{fig_system}{} (see \secreft{section:dkr_system} for a full definition
of these parameters). This system is a BEC analog of the quantum
$\delta$-kicked rotor \cite{casati_etal_1979, izrailev_shepelyanskii_1980,
saunders_etal_pra_2007, ryu_etal_prl_2006, moore_etal_prl_1995}, a paradigm
quantum-chaotic system exhibiting complex behavior as a result of the periodic
driving \cite{casati_etal_1979, izrailev_shepelyanskii_1980,
moore_etal_prl_1995}. In particular, the $\delta$-kicked rotor exhibits quantum
resonant driving frequencies, which occur whenever the kick period $T_p$ is a rational fraction of the Talbot time $T_p=4\pi$.
At these frequencies the uptake of energy from the driving
potential is greatly increased, with diffusive (linear in time) expansion of
the state in momentum space being replaced by ballistic (quadratic in time)
expansion at resonance \cite{casati_etal_1979, izrailev_shepelyanskii_1980,
saunders_etal_pra_2007}.  Such systems have previously been realized in the
context of atom-optics experiments \cite{ryu_etal_prl_2006,
moore_etal_prl_1995, oberthaler_etal_prl_1999, duffy_etal_pra_2004}. However,
the $\delta$-kicked-rotor BEC offers a route to a range of new dynamics and
phenomena, since the nonlinear effects of inter-atomic interactions introduce,
on the mean-field level, the potential for true wave chaos
\cite{shepelyansky_prl_1993, wimberger_etal_prl_2005, rebuzzini_etal_pre_2005,
gardiner_etal_pra_2000, rebuzzini_etal_pra_2007, zhang_etal_prl_2004,
liu_etal_pra_2006, reslen_etal_pra_2008, shepelyansky_prl_1993,
wimberger_etal_prl_2005, rebuzzini_etal_pre_2005, mieck_graham_jpa_2005,
monteiro_etal_prl_2009}. For experimentally realistic interaction strengths and
atom numbers, the key new dynamical features of the $\delta$-kicked-rotor BEC,
within a mean-field description, are the appearance of \textit{nonlinear}
quantum resonances (with no analog in the linear regime) and the appearance of
sharp, asymmetric cut-offs in the resonance profiles as a function of driving
period. Both these features were identified in
Ref.~\cite{monteiro_etal_prl_2009} using the zeroth-order, GPE description.

However, the beyond-mean-field effects of condensate depletion and condensate
-- non-condensate interactions can be expected to play an increasing role for
longer times in strongly non-equilibrium driven systems such as the
$\delta$-kicked-rotor BEC.  As highlighted in the introduction to this
manuscript, dealing with such dynamics within a mean-field treatment or
truncated Wigner approximation can be challenging; these challenges appear to
have precluded any treatments of the dynamics of the $\delta$-kicked-rotor BEC
using these methods.

Until recently, all explicitly beyond-mean-field studies of the
$\delta$-kicked-rotor BEC have been conducted using a first-order
number-conserving description \footnote{Note, however, that some aspects of
beyond-mean-field dynamics were inferred from analysis of the GPE in
Ref.~\cite{monteiro_etal_prl_2009}.}.  Unfortunately, the lack of a
self-consistent back-action of the non-condensate on the condensate in the
first-order number-conserving description led to unbounded growth of the
non-condensate, and unphysical results at long times when close to resonance
\cite{zhang_etal_prl_2004, liu_etal_pra_2006, reslen_etal_pra_2008}. The same
effect has also been observed in the similar
$\delta$-kicked-harmonic-oscillator BEC \cite{gardiner_etal_pra_2000,
rebuzzini_etal_pra_2007}.  However, in contrast to the first-order description,
the presence of a self-consistent back-action in the second-order
number-conserving description makes it ideal for the description of such
systems. 

\begin{figure}[t]
\centering
 \includegraphics[width=3.375in]{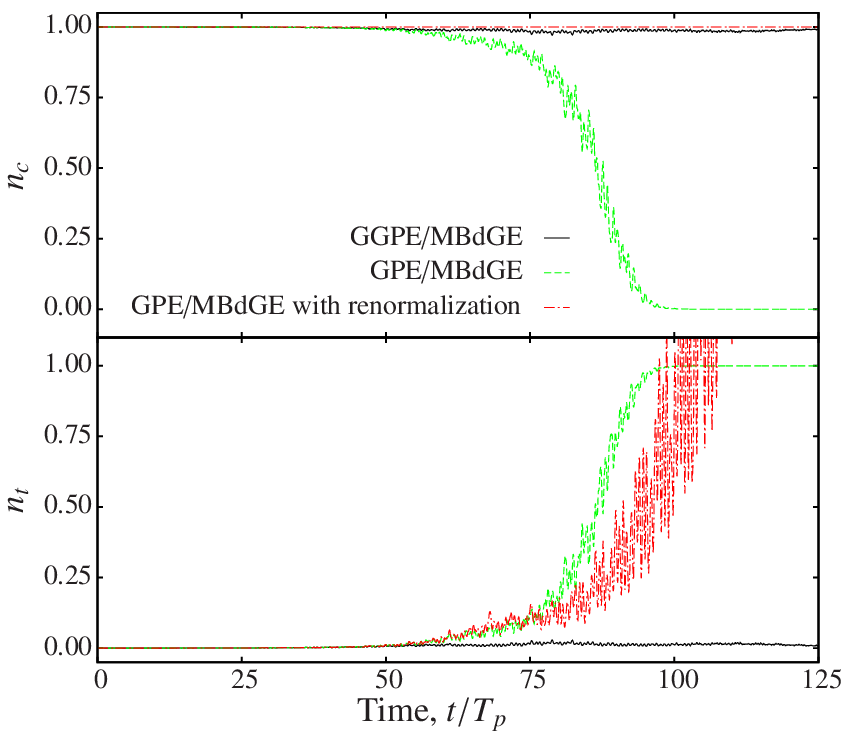}
 \caption{(color online). Zero-temperature evolution of the $\delta$-kicked-rotor BEC close to
a nonlinear resonance explored in \cite{monteiro_etal_prl_2009} (dimensionless
kick strength $\kappa = 0.5$, kick period $T_p = 6.12$, and interaction
strength $g_T = 2.5\times 10^{-4}$, and total atom number $N=10^4$ --- see
\secreft{section:dkr_system}). Beginning with a zero-temperature initial
condition, three descriptions are used to determine the dynamics: the
second-order number-conserving description [GGPE/MBdGE,
\eqnsreft{eqn:ggpe_first}{eqn:bdg_first}], first-order number-conserving
description [GPE/MBdGE, \eqnsreft{eqn:gpe}{eqn:bdg_first}], and a first-order
number-conserving description with ad-hoc renormalization. In each case we show
the evolution of the condensate fraction $n_c = N_c/N$ and the non-condensate
fraction $n_t = N_t/N$.  Note that, in the first-order description
\textit{without} renormalization, $n_c=1$ for all times.}
\label{fig:bdg}
\end{figure}

In Ref.~\cite{billam_gardiner_njp_2012} a restricted implementation of the
numerical method of \secreft{section:num_imp}, limited to the study of
zero-temperature initial conditions (and the details of which were not reported in that work), was used to demonstrate explicitly that
the self-consistent back-action of the non-condensate damps out initially rapid
growth in the non-condensate, which would continue unbounded in the first-order
description.  We illustrate this further in \figreft{fig:bdg}{}, which shows
the number-evolution obtained using the first- and second-order
number-conserving descriptions to evolve the dynamics of an initially
zero-temperature $\delta$-kicked-rotor BEC, for parameters close to a nonlinear
resonance. In addition to the evolution of the first- and second-order
descriptions, which were originally obtained in
Ref.~\cite{billam_gardiner_njp_2012}, we show the evolution of the same initial
conditions under ``renormalized'' first-order equations of motion; these are
identical to the first-order equations of motion, but with an added
renormalization of the condensate, applied by explicitly calculating $N_t$ and
setting $N_c = N - N_t$ immediately after each time step. While it does prevent
the occurrence of unphysical growth of the total particle number, this ad-hoc
renormalization procedure fails to damp out the rapid growth of the
non-condensate in a similar way to the second-order description. This
illustrates that the \textit{self-consistent} nature of the back-action of the
non-condensate in the second-order description is key to preventing unphysical
growth of the non-condensate. 

In the following Sections we outline the exact procedure for applying the
numerical method of \secreft{section:num_imp} to the $\delta$-kicked-rotor BEC.
We then use the numerical model so developed to explore the dynamics of the
$\delta$-kicked-rotor BEC for finite-temperature initial conditions for the
first time.  While no particularly novel new physics is expected beyond that
observed for zero-temperature initial conditions, this exploration nonetheless
(a) demonstrates the efficacy of the method, (b), serves as a concrete example
of how to apply the preceding formal developments to a specific problem, and
(c) allows the first \textit{quantitative} prediction of departures from the
initially zero-temperature dynamics in the finite-temperature case.Since any
future experiments on the system will necessarily operate at non-zero
temperatures, such a quantitative prediction is particularly relevant in that
regard.

\subsection{Theoretical model and initial conditions\label{section:dkr_system}}

As shown in \figreft{fig_system}{}, we consider a system described by
Hamiltonian \eqnrefp{eqn:bose_gas_hamiltonian_contact_int_recap} with a
toroidal external potential $V_{T}(\rho,z) = m\omega^2 [(\rho-R)^2+z^2]/2$
\figrefp{fig_system}{(a)}. Potentials similar to this have been experimentally
realized in, e.g., Refs.~\cite{franke-arnold_etal_oe_2007,
ramanathan_etal_prl_2011}. We assume a sufficiently strong effective trap
frequency $\omega$ that the harmonic oscillator length in this direction,
$a_r=\sqrt{\hbar/m \omega} \ll R$.  Under this assumption, a quasi-1D
approximation is justified and yields the 1D model Hamiltonian
\cite{halkyard_etal_pra_2010}
\begin{equation}
\hat{H} = \int d\x\, \foh \left[-\frac{1}{2}\frac{\partial^2}{\partial \x^2} 
+ V(\theta,t) + \frac{\is}{2} \foh \fo \right] \fo\,,
\label{eqn_many_body_hamiltonian}
\end{equation}
where we have chosen to use length units of $R$, time units of $mR^2/\hbar$,
and we have introduced the dimensionless interaction strength $\is=2a_s
R/a_r^2$.  Note that this choice of dimensionless units can be codified as
$\hbar = m \ R =1$. When working at finite-temperature, we add $k_B=1$ to this
system of units.  We restrict our analysis to the case where the system size,
$2\pi R$, is much smaller than the phase coherence length $l_\phi = 2\pi R
\exp(\sqrt{\pi N_c/2g_T})/\sqrt{4\pi g_T N_c}$; in this case the ground state
of the system can contain a true homogeneous Bose-Einstein condensate
\cite{sykes_etal_prl_2009}. We do not consider the alternative case of a
quasicondensate ($l_\phi \lesssim 2\pi R$) \cite{kagan_etal_jetp_1987,
mora_castin_pra_2003}.  

As in previous studies of the $\delta$-kicked-rotor BEC
\cite{zhang_etal_prl_2004, liu_etal_pra_2006, reslen_etal_pra_2008,
monteiro_etal_prl_2009}, we model the driving potential as a train of
$\delta$-kicks
\begin{equation}
V(\x,t) = \kappa \cos(\x)\sum_{n=0}^{\infty} \delta(t-nT_p)\,.
\end{equation}
Here the dimensionless kicking period $T_p$ is defined in terms of the real
kick period $T_p^{(\mathrm{real})}$ as $T_p = \hbar
T_p^{(\mathrm{real})}/mR^2$, and the dimensionless kick strength $\kappa$ is
dependent on the real strength of the driving potential, and its duration
$t_p$. Such a driving potential may be approximated in experiment using, e.g.,
short pulses of off-resonant laser light \cite{saunders_etal_pra_2007,
ryu_etal_prl_2006, moore_etal_prl_1995}. Further details of the physical
aspects of the system can be found in Refs.~\cite{billam_gardiner_njp_2012,
monteiro_etal_prl_2009}.

The equations of motion for the $\delta$-kicked-rotor BEC are identical to those of
\secreft{section:eoms} with the replacement $\vx \rightarrow \x$. The dynamical
and thermal equilibrium state of the system consists of a uniform condensate
mode, $\psi(\x) = \sqrt{N_c/2\pi}$, accompanied by a thermal population of
Bogoliubov quasiparticle excitations. Both at $T=0$ and for $T>0$, the
stationary initial Bogoliubov modes associated with the uniform condensate are
given by \cite{billam_gardiner_njp_2012}
\begin{equation}
\twovec{u_k(\x)}{v_k(\x)} = \frac{1}{2}\twovec{C_k + C_k^{-1}}{C_k - C_k^{-1}}\frac{e^{ik\x}}{\sqrt{2\pi}},
\label{eqn_init_qp}
\end{equation}
where 
\begin{equation}
C_k = C_{-k} =  \left(\frac{k^2}{k^2+4\lambda_0^{(2)}}\right)^{1/4}\,,
\end{equation}
and, for the uniform initial condensate, $\lambda_0^{(2)} = \is N_{c}/2\pi$. It is
important to note that in this Section we have replaced the completely generic
quasiparticle index $k = 1,2,\ldots$ used in \secreft{section:num_imp} with the
quasiparticle momentum index $k = \ldots,-2,-1,1,2,\ldots$, as this
considerably simplifies the notation.  Consequently, in our numerical treatment
we must also replace the generic quasiparticle cut-off momentum $M$ with an
explicit maximum quasiparticle momentum index $k_\mathrm{max}$. The expressions
appearing in \secreft{section:num_imp} can be reformulated in this new notation
simply by making the replacements 
\begin{align}
\sum_{k=1}^M &\rightarrow \sideset{}{'}\sum_{k =-k_\mathrm{max}}^{k_\mathrm{max}}\,,\\
M &\rightarrow 2k_\mathrm{max}\,,
\end{align}
where $\sum^\prime$ indicates a summation omitting the term $k=0$ (which here
corresponds to the condensate mode).

The energy of a quasiparticle of momentum $k$ is given by
\begin{equation}
\epsilon_k = \sqrt{\frac{k^4}{4}+k^2 \lambda_0^{(2)}}\,.
\end{equation}
Hence, the non-condensate population can be determined from 
\begin{multline}
\label{eqn:nt_analytic}
N_t = \int_0^{2\pi} d\x \sideset{}{'}\sum_{k =-k_\mathrm{max}}^{k_\mathrm{max}} \left[N_k |u_k(\x)|^2 + (N_k+1) |v_k(\x)|^2\right]\\
 =\frac{1}{4}\sideset{}{'}\sum_{k =-k_\mathrm{max}}^{k_\mathrm{max}} \left[N_k \left(C_k+C_{-k}^{-1}\right)^2 + (N_k+1) \left(C_k-C_{-k}^{-1}\right)^2\right]\,,
\end{multline}
where the populations $N_k$ are obtained from the energies $\epsilon_k$ using
the appropriate \textit{thermal} Bose distribution \eqnrefp{eqn:pops}. Note
that in our system of dimensionless variables, the dimensionless temperature
$T$ used to obtain populations corresponds to a real temperature through the
relation $T_\mathrm{real} = T \hbar^2/m R^2 k_B$. Our procedure for finding a
self-consistent initial condition thus represents an extension of the method
used in Ref.~\cite{billam_gardiner_njp_2012}, which allows us to simulate
finite temperature dynamics. 

To obtain a specific initial condition, we initially set $N_c = N$ and then:
(a) calculate the coefficients $C_k$ up to the momentum cut-off
$k_\mathrm{max}$; (b) calculate $N_t$ using \eqnreft{eqn:nt_analytic}; (c)
renormalize the total number of atoms by reducing the condensate population to
$N_c = N - N_t$. Steps (a--c) are then repeated until $N_t$ converges to within
$10^{-6}$ of its final value. We choose $k_\mathrm{max}$ sufficiently large
that $N_{k_\mathrm{max}} < 10^{-2}$ (in addition to confirming that the
subsequent dynamics have converged as a function of $k_\mathrm{max}$); this
means that $k_\mathrm{max}$ increases substantially with temperature.  However,
for the temperatures we consider (up to $T=300$) $k_\mathrm{max}=64$ is
sufficient, and the effect of the finite-size correction $\mu - \lambda_0^{(2)}$ is
negligible. Finally, we numerically evaluate the GGPE eigenvalue
$\bar{\lambda}_2^{(2)}$ for the obtained initial equilibrium condition. Evolution of
the resulting initial condition is accomplished using the numerical method of
\secreft{section:num_imp}. We choose time-steps which exactly divide the
dimensionless kick period $T_p$, such that the effect of a kick can be
accomplished via the instantaneous transformation
\begin{align}
\psi(\x) &\rightarrow e^{-i\kappa\cos\x}\psi(\x)\,,\\
u_k(\x) &\rightarrow e^{-i\kappa\cos\x}u_k(\x)\,,\\
v_k(\x) &\rightarrow e^{i\kappa\cos\x}v_k(\x)\,,
\end{align}
which is applied at the exact instant of each kick.  The resulting evolution is
then checked for convergence in the number of grid points, quasi-particle
momentum cut-off, and time-step.

\subsection{Finite-temperature dynamics\label{section:dkr_ft}}

We now explore the dynamics of the $\delta$-kicked rotor for a range of finite
temperature equilibrium initial conditions. In addition to tracking the
condensate fraction $n_c = N_c / N$ and non-condensate fraction $n_t = N_t / N$
--- variables which the number-conserving description gives immediate access to
--- we also introduce measures to track the presence of many-body,
beyond-mean-field effects in the system.  In
Ref.~\cite{billam_gardiner_njp_2012}, the quantity
\begin{equation}
\g = \iint d\x\, d\x^\prime\, g_{1}(\x,\x^\prime) g_{1}(\x^\prime,\x),
\label{eqn_g_def}
\end{equation}
which represents a spatial average of the first-order correlation function
$g_{1}(\x,\x^\prime) = \avg{\hat{\Psi}^\dagger(\x^\prime)\hat{\Psi}(\x)}/N$,
was introduced for this purpose.  In terms of the single-particle density
matrix $\rho(\x,\x^\prime)$, $\g$ can be defined as
\begin{equation}
\label{eqn:g_def_spdm}
\g = \frac{\mathop{\mbox{Tr}}\left\{\rho(\x,\x^\prime)^2 \right\}}{N^2}\,.
\end{equation}
To an order consistent with the rest of the second-order number-conserving
description, the single particle density matrix $\rho(\x,\x^\prime)$ is given
by
\begin{multline}
\label{eqn:num_density_mat}
\rho(\x,\x^\prime) = \psi^\ast(\x^\prime) \psi(\x) + \sideset{}{'}\sum_{k =-k_\mathrm{max}}^{k_\mathrm{max}} \left[N_k u_k(\x) u_k^\ast(\x') \right]\\
+ \sideset{}{'}\sum_{k =-k_\mathrm{max}}^{k_\mathrm{max}} \left[(N_k +1) v_k(\x') v_k^\ast(\x) \right]\,.
\end{multline}
The coherence measure $\g$ is thus analogous to the purity of a state in a
single-particle system, defined by $\mathrm{Tr}[\varrho^2]/(\mathrm{Tr}[\varrho])^2$ (where
$\varrho$ is the full density matrix for the single-particle system). Using
this measure an entirely condensed state (zero non-condensate fraction) is akin
to a pure state of a single-particle system, and has $\g=1$. The presence of
non-condensate --- either in the form of the zero-temperature quantum
depletion, or in the form of thermally excited quasiparticles --- guarantees
that $\g<1$. This is akin to a mixed state of a single-particle system.
Physically, this can be interpreted as indicating the presence of high-order
(many-body) correlations between the condensate and non-condensate; these are
effectively ``traced out'' when one uses the number-conserving description to
compute a single-particle density matrix, resulting in $\rho(\x,\x^\prime)$
appearing ``mixed''.

In this paper we introduce another measure which probes the underlying
many-body correlations: the single-particle von Neumann entropy.  This again is
defined in terms of the single-particle density matrix, and is given by
\begin{equation}
\label{eqn:s_sp_def}
S_\mathrm{sp} = -\mathop{\mbox{Tr}}\left\{\frac{\rho}{N} \ln \left( \frac{\rho}{N} \right) \right\}\,.
\end{equation}
In contrast to $\g$, which is unity for a pure condensate, $S_\mathrm{sp} = 0$
for a pure condensate, and $S_\mathrm{sp}>0$ otherwise. Much like $\g$,
however, $S_\mathrm{sp}$ gives an indication of the magnitude of the underlying
many-body correlations and the degree of entanglement present in the system. It
should be noted that as a direct measure of entanglement in a many-body system,
$S_\mathrm{sp}$ is imperfect, as a proportion of its magnitude can be due to
Bose-symmetry rather than genuine entanglement (i.e., entanglement between
particles not arising as a direct consequence of the Bose-symmetry of the
wavefunction \cite{amico_etal_rmp_2008, ghirardi_marinatto_os_2005,
vedral_cejp_2003}).  However, its use as an \textit{indicator} for such
entanglement, as we use it here, is well-established \cite{amico_etal_rmp_2008,
vardi_anglin_prl_2001, sowinski_etal_pra_2010}.

\begin{figure}[t]
\centering
 \includegraphics[width=3.375in]{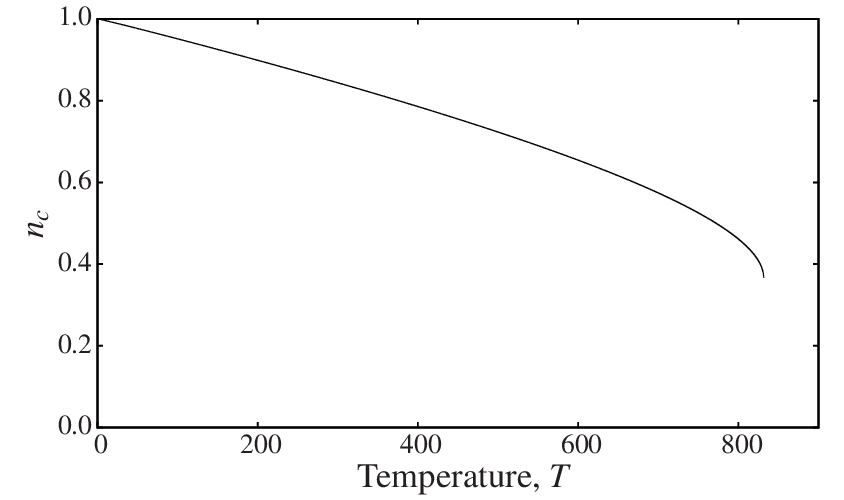}
 \caption{Condensate fraction, $n_c$, in the equilibrium ground state of the
$\delta$-kicked-rotor BEC (parameters $g_T = 2.5\times 10^{-4}$, $N=10^4$) as a
function of the dimensionless temperature, $T$, calculated within the
second-order number-conserving description. The point above which convergent
solutions cannot be found gives an estimate of the condensation transition
temperature $T_c \approx 832$ for this system. However, our second-order
treatment is valid only for temperatures such that the small parameter
$(1-n_c)/n_c \ll 1$.}
\label{figure_depletion}
\end{figure}

\begin{figure}[t]
\centering
 \includegraphics[width=3.375in]{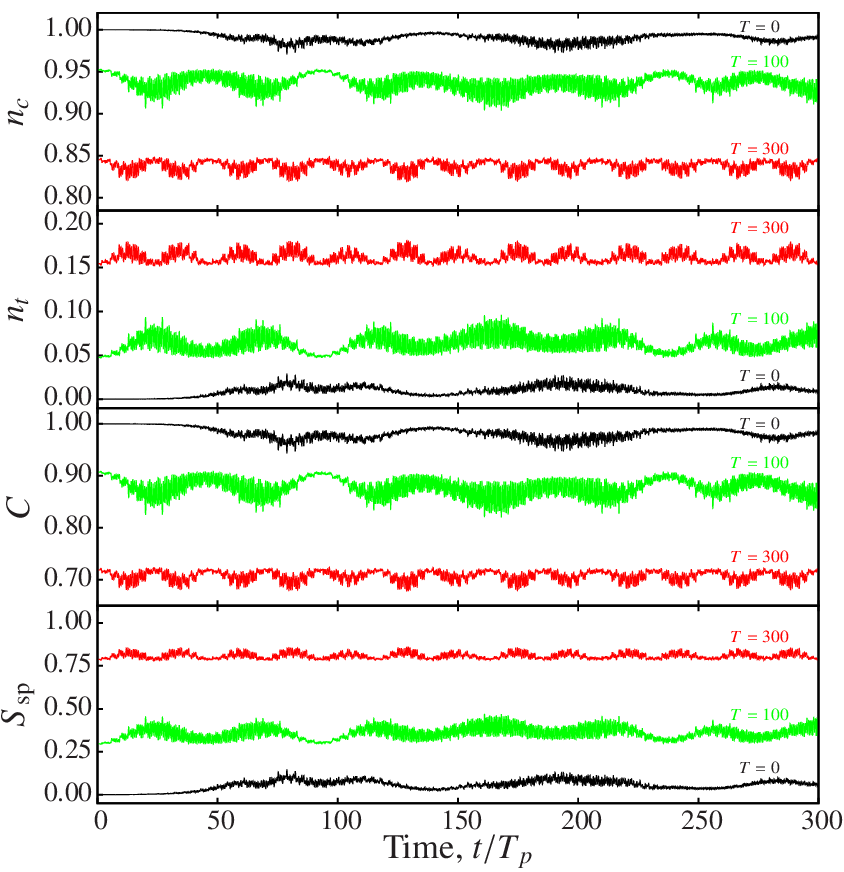}
 \caption{(color online). Finite-temperature evolution of the $\delta$-kicked-rotor BEC close
to a nonlinear resonance explored in \cite{monteiro_etal_prl_2009} (parameters
$g_T = 2.5\times 10^{-4}$, $N=10^4$, $T_p=6.12$, $\kappa=0.5$).  Condensate and
non-condensate fractions $n_c$ and $n_t$, coherence measure $C$, and the
single-particle von Neumann entropy, $S_\mathrm{sp}$ are shown for initial
temperatures $T=0$, $T=100$, and $T=300$.}
\label{figure_Numbers}
\end{figure}

To explore the finite-temperature evolution of the $\delta$-kicked-rotor BEC,
we initially consider specific parameters ($T_p = 6.12$, $\kappa=0.5$,
$g_T=2.5\times 10^{-4}$, and $N=10^4$) which are close to a nonlinear
resonance; these parameters were previously subject to detailed analysis ---
using the GPE, the first-order number-conserving description, and the
second-order number-conserving description --- in
Refs.~\cite{monteiro_etal_prl_2009} and \cite{billam_gardiner_njp_2012}.
\figreftfull{figure_Numbers}{} shows the evolution of the
$\delta$-kicked-rotor BEC, for these parameters, resulting from three different
equilibrium initial conditions with temperatures $T=0$, $T=100$, and $T=300$.
These dimensionless temperatures should be considered in the context of
\figreft{figure_depletion}{}, which shows the condensate fraction in the
equilibrium ground state of this system (calculated as described in
\secreft{section:dkr_system}) within our second-order, number-conserving
treatment. Within this treatment $T_c\approx 832$ is the effective critical
temperature, above which it is no longer possible to converge a self-consistent
equilibrium initial condition. Note, however, that we choose temperatures
sufficiently low that $n_c > 0.8$ such that $\tilde{\Lambda}(\vx)$ remains a
small parameter.

For each initial condition the condensate fraction $n_c$, non-condensate
fraction $n_t$, coherence measure $\g$, and single-particle von Neumann entropy
$S_\mathrm{sp}$ are shown in \figreft{figure_Numbers}{}. In each case, the
coherence measure and single-particle von Neumann entropy follow the
non-condensate fraction very closely. This indicates that depletion of the
condensate is the dominant factor in creating many-body correlations and
entanglement in this system.

\begin{figure}[t]
\centering
 \includegraphics[width=3.375in]{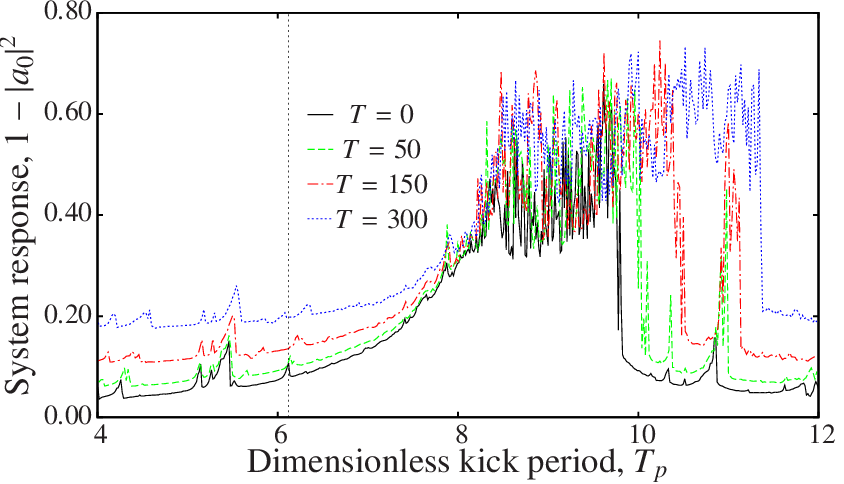}
 \caption{(color online). Finite-temperature shift of quantum resonances in the
$\delta$-kicked-rotor BEC.  Parameters $g_T = 2.5\times10^{-4}$, $N = 2.5$, and
$\kappa=0.5$ have been chosen such that $T_p=6.12$ (vertical dashed line)
corresponds to the nonlinear resonance explored in
\cite{monteiro_etal_prl_2009,billam_gardiner_njp_2012} and
\figreft{figure_Numbers}. In particular, the finite-temperature shift in the
location of this nonlinear resonance explains the non-resonant behavior seen
at finite temperature in \figreft{figure_Numbers}. Interestingly, the extremely
sharp cut-offs observed in the GPE \cite{monteiro_etal_prl_2009} and at
zero-temperature in the GGPE \cite{billam_gardiner_njp_2012} are preserved at
$T=300$.}
\label{figure_FiniteTempResonances}
\end{figure}

The most noticeable difference between the temperatures
is in the initial non-condensate fraction (and consequently in the coherence
and entropy, since these generally scale with $n_t$ as explained above). This
increase is to be expected, as the number of thermally excited quasiparticles
increases as a function of temperature. 
The larger thermal population at higher temperatures also has the effect of
significantly depleting the condensate, and hence reducing the effective
nonlinearity, as given by the product $g_T N$.  At zero temperature, the
resonant kicking period is known to shift downwards as a function of the
product $g_T N$, which represents the effective nonlinearity
\cite{monteiro_etal_prl_2009, billam_gardiner_njp_2012}. Hence, the presence of
a thermal background, which effectively reduces $g_T N$, produces an upwards
shift in the resonant kicking period. That this is the case is confirmed
numerically in \figreft{figure_FiniteTempResonances}{}, in which we plot the
overall response of the system --- measured by the fractional population of all
momentum modes higher than $k=0$ among all atoms averaged over 100 kick periods
(see \cite{billam_gardiner_njp_2012}) --- as a function of kicking period $T_p$
and temperature. Another interesting feature revealed by
\figreft{figure_FiniteTempResonances}{} is that the sharp cut-off at the upper
limit of the shifted Talbot-time resonance (the large resonant area to the
right of the graphs) first observed in \cite{monteiro_etal_prl_2009} is
qualitatively preserved at finite temperatures. This is interesting as these
features are not present in the non-interacting limit $g_T N_c \rightarrow 0$,
and one might therefore reasonably expect that the decrease in condensate
population (and associated decrease in coherence) resulting from the increasing
population of quasiparticles at finite-temperature would gradually soften these
features. However, we observe that the features remain sharp even with
$\sim$20\% depletion of the condensate.

While the dynamics of the system at the finite temperatures accessible with the
second-order number-conserving description show no major qualitative
differences to the initially zero-temperature case, the quantitative shifts in
the positions of the resonances would be particularly relevant in the context
of experiments aimed at studying nonlinear resonances in the
$\delta$-kicked-rotor BEC. For example, in an experiment with similar
parameters to Ref.~\cite{ramanathan_etal_prl_2011}, the unit of temperature
defined above takes a value $\sim 10$~pK. Consequently, a future experiment
with a temperature value in the range $1$--$10$~nK would have dimensionless
temperature $100 \lesssim T \lesssim 1000$ in the units used here. Our results
show that the nonlinear resonances in such an experiment would thus be
considerably shifted away from their zero-temperature values, and provide a
quantitative prediction of this shift which could be experimentally tested. The
preserved sharpness of the resonance cut-offs would help facilitate such a
measurement.

\section{Conclusions\label{section:conc}}

We have described in detail a numerical method for evolving the
integro-differential equations of motion of the second-order number-conserving
description of Gardiner and Morgan \cite{gardiner_morgan_pra_2007}. This
numerical method explicitly includes problematic nonlinear, non-local terms
which are necessary in order to preserve orthogonality between condensate and
non-condensate in this description. Our method provides a fully time-dependent,
self-consistent treatment of number-dynamics and condensate -- non-condensate
interactions in a finite-temperature BEC, and provides an excellent framework
with which to study the non-equilibrium dynamics of driven BEC systems at low
temperatures.  We have used this numerical method to systematically study a
prototypical example of such a system, the $\delta$-kicked-rotor BEC, at finite
temperature. While the qualitative features of the zero-temperature dynamics in
this system are generally preserved at finite temperatures, our treatment
nonetheless provides the first prediction of a quantitative shift in resonance
frequencies at higher temperatures.  This shift would be relevant for, and
could feasibly be verified by, future experiments studying this system.

\begin{acknowledgments}
We thank A. S. Bradley, S. Fishman, T. S. Monteiro, M. D. Lee, and H. Veksler
for enlightening discussions, and acknowledge support from the UK Engineering
and Physical Sciences Research Council (Grant No.  EP/G056781/1), the Marsden
Fund of New Zealand (contract UOO162) and the Royal Society of New Zealand
(contract UOO004) (T.P.B.), Marie Curie Fellowship NUM2BEC (No. 300285) (P.M.),
and the Jack Dodd Centre and The Royal Society of London (Grant IE110202)
(S.A.G.). 
\end{acknowledgments}

\appendix
\section{Matrix form of the evolution operator\label{section:app}}

\subsection{Determination of eigenvalues}
We wish to find a general analytic expression for the matrix form of
$e^{-i\Gamma_\mathrm{N}(\vx,t)\delta t / 2\hbar}$ at a given position, $\vx$
and time $t$, where $\Gamma_\mathrm{N}(\vx,t)$ is defined by 
\begin{widetext}
\begin{equation}
\Gamma_\mathrm{N}(\vx,t) =
\left( \begin{array}{cccccccccc}
H_{\rm N}(\vx,t) + B(\vx,t) & (N_1 +1) A_1(\vx,t) & (N_2 +1) A_2(\vx,t) &\cdots & (N_M +1) A_M(\vx,t) & N_1 A_1^\ast(\vx,t) & N_2 A_2^\ast(\vx,t) & \cdots & N_M A_M^\ast(\vx,t) \\
A_1^\ast(\vx,t) & H_{\rm N}(\vx,t) & 0 & \cdots &0 & 0 & 0 & \cdots & 0 \\
A_2^\ast(\vx,t) & 0 & H_{\rm N}(\vx,t) & \cdots &0 & 0 & 0 & \cdots & 0 \\
\vdots & \vdots &\vdots & \ddots & \vdots & \vdots & \vdots & \cdots & \vdots \\
A_M^\ast(\vx,t) & 0 & 0 & \cdots & H_{\rm N}(\vx,t) & 0 & 0 & \cdots & 0 \\
A_1(\vx,t) & 0 & 0 & \cdots &0 & H_{\rm N}(\vx,t) & 0 & \cdots & 0 \\
A_2(\vx,t) & 0 & 0 & \cdots &0 & 0 & H_{\rm N}(\vx,t) & \cdots & 0 \\
\vdots & \vdots & \vdots & \vdots & \vdots & \vdots & \vdots & \ddots & \vdots \\
A_M(\vx,t) & 0 & 0 & \cdots &0 & 0 & 0 & \cdots & H_{\rm N}(\vx,t) \\
\end{array}\right)\,,
\end{equation}
for 
\begin{equation}
H_\mathrm{N}(\vx,t) = V(\vx,t) + U_0|\psi(\vx,t)|^2\,,
\end{equation}
and
\begin{equation}
B(\vx,t) = U_0 \tilde{n}(\vx,\vx,t)- U_0 \frac{|\psi(\vx,t)|^2}{N_c} - \lambda^\prime\,.
\end{equation}
We will show that this exponentiation of $\Gamma_\mathrm{N}(\vx,t)$, at a given
position and time, can be accomplished in a closed analytic form for arbitrary
$M$.  From this point forward, we adopt the clarifying convention of omitting
the explicit position and time arguments $\vx$ and $t$: all quantities
appearing in our treatment should be interpreted as the complex,
\textit{scalar} values obtained by evaluating the corresponding $\vx$- and
$t$-dependent functions at a particular position and time.

We proceed by identifying all eigenvalues of the matrix $\Gamma_\mathrm{N}$
through the characteristic equation
\begin{equation}
\mathrm{det} \left( \Gamma_\mathrm{N} - \xi I \right) = 0\,,
\end{equation}
where $I$ is the $(2M+1)\times(2M+1)$ identity matrix.  Expanding the
determinant over minors of the first row yields
\begin{multline}
\mathrm{det} \left( \Gamma_\mathrm{N} - \xi I \right) =\left( H_{\rm N} + B -\xi \right) \left(H_{\rm N}-\xi\right) I
- (N_1 + 1) A_1 \times
\left|\begin{array}{ccccccccc}
A_1^\ast & 0 & \cdots &0 & 0 & 0 & \cdots & 0 \\
A_2^\ast & H_{\rm N}-\xi & \cdots &0 & 0 & 0 & \cdots & 0 \\
\vdots &\vdots & \ddots & \vdots & \vdots & \vdots & \cdots & \vdots \\
A_M^\ast & 0 & \cdots & H_{\rm N}-\xi & 0 & 0 & \cdots & 0 \\
A_1 & 0 & \cdots &0 & H_{\rm N}-\xi & 0 & \cdots & 0 \\
A_2 & 0 & \cdots &0 & 0 & H_{\rm N}-\xi & \cdots & 0 \\
\vdots & \vdots & \vdots & \vdots & \vdots & \vdots & \ddots & \vdots \\
A_M & 0 & \cdots &0 & 0 & 0 & \cdots & H_{\rm N}-\xi \\
\end{array}\right|
\\+ (N_2 +1) A_2 \times
\left|\begin{array}{ccccccccc}
A_1^\ast & H_{\rm N}-\xi & \cdots &0 & 0 & 0 & \cdots & 0 \\
A_2^\ast & 0 & \cdots &0 & 0 & 0 & \cdots & 0 \\
\vdots & \vdots & \ddots & \vdots & \vdots & \vdots & \cdots & \vdots \\
A_M^\ast & 0 & \cdots & H_{\rm N}-\xi & 0 & 0 & \cdots & 0 \\
A_1 & 0 & \cdots &0 & H_{\rm N}-\xi & 0 & \cdots & 0 \\
A_2 & 0 & \cdots &0 & 0 & H_{\rm N}-\xi & \cdots & 0 \\
\vdots & \vdots & \vdots & \vdots & \vdots & \vdots & \ddots & \vdots \\
A_M & 0 & \cdots &0 & 0 & 0 & \cdots & H_{\rm N}-\xi \\
\end{array}\right|
 -\ldots 
\label{eqn:detexp}
\end{multline}
where the choice of $+$ and $-$ signs in later terms is dependent on the parity
of $M$.  The determinant of the first minor is trivially $(H_{\rm
N}-\xi)^{2M}$, and the determinant of each subsequent minor is easily computed
by further expanding in minors along the row containing no entry of the form
$H_{\rm N}-\xi$. Carefully keeping track of signs, this yields
\begin{multline}
\mathrm{det} \left(\Gamma_\mathrm{N} - \xi I \right) = \left(H_{\rm N}+B-\xi \right)\left(H_{\rm N}-\xi\right)^{2M}
+ \sum_{k=1}^M (-1)^k (N_k +1) A_k (-1)^{k+1} A_k^\ast \left(H_{\rm N}-\xi\right)^{2M-1} \\
+ \sum_{k=1}^M (-1)^{M+k} N_k A_k^\ast (-1)^{M+k+1} A_k \left(H_{\rm N}-\xi\right)^{2M-1}\,,
\end{multline}
which simplifies to
\begin{equation}
\mathrm{det} \left(\Gamma_\mathrm{N} - \xi I \right) =  \left[ \left(H_{\rm N}+B-\xi \right)\left(H_{\rm N}-\xi\right) - \Sigma \right] \left(H_{\rm N}-\xi\right)^{2M-1} \,,
\label{eqn:det_final}
\end{equation}
where
\begin{equation}
\Sigma = \sum_{k=1}^M (2N_k +1) |A_k|^2\,.
\end{equation}
Note that \eqnreft{eqn:det_final} holds regardless of the parity of $M$.
Consequently, $2M-1$ eigenvalues of $\Gamma_\mathrm{N}$ are degenerate, having
value $H_{\rm N}$. The remaining two eigenvalues $\xi$ solve the quadratic
equation
\begin{equation}
\left[H_{\rm N}-\xi \right]^2 + B\left[H_{\rm N}-\xi\right] - \Sigma = 0 \,,
\end{equation}
and hence are given by
\begin{equation}
\xi = H_{\rm N} + \frac{B}{2} \pm \sqrt{\left(\frac{B}{2}\right)^2 + \Sigma }\,.
\end{equation}

\subsection{Determination of eigenvectors}
A set of linearly independent eigenvectors $\zeta$ associated with the $2M-1$
degenerate eigenvalues $\xi = H_{\rm N}$ must be obtained by solving the linear
equations
\begin{equation}
\label{eqn:lineigvec}
\left( \Gamma_\mathrm{N}-H_{\rm N}I \right) \zeta = 0\,.
\end{equation}
Such a set can be found by setting $\psi = 0$ in Eqs.~\ref{eqn:lineigvec}, reducing the linear equations to
\begin{equation}
\sum_{k=1}^M \left[ (N_k +1) A_k v_k^\ast + N_k A_k^\ast u_k \right] = 0\,.
\end{equation}
The required linearly independent solutions are given by
\begin{align}
v_1^\ast &= (N_k + 1)A_k\,,\nonumber\\
v_k^\ast &= -\delta_{kj} (N_1 + 1) A_1\,,\nonumber\\
u_k &= 0\,,
\end{align}
where $j \in \{2 \ldots M\}$, and
\begin{align}
v_1^\ast &= N_k A_k^\ast\,,\nonumber\\
v_k^\ast &= 0 \,, \nonumber\\
u_k &= -\delta_{kj} (N_1 + 1) A_1\,,
\end{align}
where $j \in \{1 \ldots M\}$.

For the remaining two eigenvalues one has to solve the linear equations,
\begin{equation}
\left\{ \Gamma_\mathrm{N}-\left[H_{\rm N}+\left(\frac{B}{2}\pm \sqrt{\left(\frac{B}{2}\right)^2 + \Sigma }\right) \right] I \right\} \zeta = 0\,,
\end{equation}
which can be expressed as the system
\begin{gather}
A_k^\ast \psi - \left(\frac{B}{2} \pm \sqrt{\left(\frac{B}{2}\right)^2 + \Sigma } \right) v_k^\ast = 0\,,\\
A_k \psi - \left(\frac{B}{2} \pm \sqrt{\left(\frac{B}{2}\right)^2 + \Sigma } \right) u_k = 0\,,\\
\left(\frac{B}{2} \mp \sqrt{\left(\frac{B}{2}\right)^2 + \Sigma} \right) \psi
+ \sum_{k=1}^M (N_k +1) A_k v_k^\ast + \sum_{k=1}^M N_k A_k^\ast u_k =0\,.
\end{gather}
Rearranging the first two equations to obtain
\begin{align}
v_k^\ast &= \frac{A_k^\ast}{\frac{B}{2} \pm \sqrt{\left(\frac{B}{2}\right)^2 + \Sigma }} \psi\,,\\
u_k &= \frac{A_k}{\frac{B}{2} \pm \sqrt{\left(\frac{B}{2}\right)^2 + \Sigma }} \psi\,,
\end{align}
provides a solution for arbitrary $\psi$, since substituting the above
expressions into the third equation yields
\begin{equation}
\left[
\left(\frac{B}{2} \mp \sqrt{\left(\frac{B}{2}\right)^2 + \Sigma} \right) 
\left(\frac{B}{2} \pm \sqrt{\left(\frac{B}{2}\right)^2 + \Sigma} \right) + \Sigma \right] \psi = 0\,,
\end{equation}
and hence
\begin{equation}
\left\{ \left(\frac{B}{2}\right)^2 - \left[\left(\frac{B}{2}\right)^2 + \Sigma \right] + \Sigma \right\} \psi = 0\,,
\end{equation}
which is indeed satisfied for all $\psi$. 

In the absence of any overriding scheme for normalizing the eigenvectors, a
practically useful approach is to eliminate denominators in the transformation
matrix $P$ which has the eigenvectors as its columns. Choosing such a
normalization, $P$ is given by:
\begin{equation}
P =
\left( \begin{array}{cccccccccc}
B_+ & B_- & 0 &\cdots & 0 & 0 & 0 & \cdots & 0 \\
A_1^\ast & A_1^\ast & (N_2+1)A_2 & \cdots &(N_M+1)A_M & N_1A_1^\ast & N_2A_2^\ast & \cdots & N_MA_M^\ast \\
A_2^\ast & A_2^\ast & -(N_1 + 1) A_1 & \cdots &0 & 0 & 0 & \cdots & 0 \\
\vdots & \vdots &\vdots & \ddots & \vdots & \vdots & \vdots & \cdots & \vdots \\
A_M^\ast & A_M^\ast & 0 & \cdots & -(N_1 + 1) A_1 & 0 & 0 & \cdots & 0 \\
A_1 & A_1 & 0 & \cdots &0 & -(N_1 + 1)A_1 & 0 & \cdots & 0 \\
A_2 & A_2 & 0 & \cdots &0 & 0 & -(N_1 +1)A_1 & \cdots & 0 \\
\vdots & \vdots & \vdots & \vdots & \vdots & \vdots & \vdots & \ddots & \vdots \\
A_M & A_M & 0 & \cdots &0 & 0 & 0 & \cdots & -(N_1 +1)A_1 \\
\end{array}\right)\,,
\end{equation}
where
\begin{equation}
B_\pm = \frac{B}{2} \pm \sqrt{\left(\frac{B}{2}\right)^2 + \Sigma }\,.
\end{equation}

\subsection{Matrix Exponentiation}
In order to exponentiate the matrix $\Gamma_\mathrm{N}$, one also requires the
inverse, $P^{-1}$, of the transformation matrix $P$. Note that, since
$\Gamma_\mathrm{N}$ is not Hermitian, $P$ is not unitary, and $P^{-1} \ne
P^\dagger$. However, $P$ can nonetheless be inverted, for non-zero $A_k$, by a
lengthy series of elementary row operations. This yields
\begin{equation}
P^{-1} =\frac{1}{\Sigma}
\left( \begin{array}{cccccccccc}
\frac{\Sigma}{B_+-B_-} & C_- (N_1 +1)A_1 & C_- (N_2+1)A_2 &\cdots & C_-(N_M+1)A_M & C_-N_1A_1^\ast & C_-N_2A_2^\ast & \cdots & C_-N_MA_M^\ast \\
\frac{-\Sigma}{B_+-B_-} & C_+ (N_1 +1)A_1 & C_+ (N_2+1)A_2 &\cdots & C_+(N_M+1)A_M & C_+N_1A_1^\ast & C_+N_2A_2^\ast & \cdots & C_+N_MA_M^\ast \\
0 & \frac{A_2^\ast (N_1+1)A_1}{(N_1+1)A_1} & \frac{A_2^\ast(N_2+1)A_2 - \Sigma}{(N_1 + 1) A_1} & \cdots & \frac{A_2^\ast(N_M+1)A_M}{(N_1+1)A_1} & \frac{A_2^\ast N_1 A_1^\ast}{(N_1+1)A_1} & \frac{A_2^\ast N_2 A_2^\ast}{(N_1+1)A_1} & \cdots & \frac{A_2^\ast N_M A_M^\ast}{(N_1+1)A_1} \\
\vdots & \vdots &\vdots & \ddots & \vdots & \vdots & \vdots & \cdots & \vdots \\
0 & \frac{A_M^\ast (N_1+1)A_1}{(N_1+1)A_1} & \frac{A_M^\ast(N_2+1)A_2}{(N_1 + 1) A_1} & \cdots & \frac{A_M^\ast(N_M+1)A_M - \Sigma}{(N_1+1)A_1} & \frac{A_M^\ast N_1 A_1^\ast}{(N_1+1)A_1} & \frac{A_M^\ast N_2 A_2^\ast}{(N_1+1)A_1} & \cdots & \frac{A_M^\ast N_M A_M^\ast}{(N_1+1)A_1} \\
0 & \frac{A_1 (N_1+1)A_1}{(N_1+1)A_1} & \frac{A_1(N_2+1)A_2}{(N_1 + 1) A_1} & \cdots & \frac{A_1(N_M+1)A_M}{(N_1+1)A_1} & \frac{A_1 N_1 A_1^\ast - \Sigma}{(N_1+1)A_1} & \frac{A_1 N_2 A_2^\ast}{(N_1+1)A_1} & \cdots & \frac{A_1 N_M A_M^\ast}{(N_1+1)A_1} \\
0 & \frac{A_2 (N_1+1)A_1}{(N_1+1)A_1} & \frac{A_2(N_2+1)A_2}{(N_1 + 1) A_1} & \cdots & \frac{A_2(N_M+1)A_M}{(N_1+1)A_1} & \frac{A_2 N_1 A_1^\ast}{(N_1+1)A_1} & \frac{A_2 N_2 A_2^\ast - \Sigma}{(N_1+1)A_1} & \cdots & \frac{A_2 N_M A_M^\ast}{(N_1+1)A_1} \\
\vdots & \vdots & \vdots & \vdots & \vdots & \vdots & \vdots & \ddots & \vdots \\
0 & \frac{A_M (N_1+1)A_1}{(N_1+1)A_1} & \frac{A_M(N_2+1)A_2}{(N_1 + 1) A_1} & \cdots & \frac{A_M(N_M+1)A_M}{(N_1+1)A_1} & \frac{A_M N_1 A_1^\ast}{(N_1+1)A_1} & \frac{A_M N_2 A_2^\ast}{(N_1+1)A_1} & \cdots & \frac{A_M N_M A_M^\ast-\Sigma}{(N_1+1)A_1} \\
\end{array}\right)\,,
\end{equation}
where
\begin{equation}
C_\pm = \frac{1}{2} \pm \frac{B}{4\sqrt{\left(\frac{B}{2}\right)^2 + \Sigma}}\,.
\end{equation}

Now we have $\Gamma_\mathrm{N} = P D P^{-1}$, where $D$ is the diagonal matrix
of the eigenvalues of $\Gamma_\mathrm{N}$. Consequently, we can compute the
matrix exponential
\begin{equation}
e^{-i\Gamma_\mathrm{N} \delta t/2\hbar} = P e^{-iD\delta t /2 \hbar} P^{-1}\,,
\end{equation}
using the results we have obtained up to this point. Using the identity,
\begin{equation}
C_- B_+ = -C_+ B_- = \frac{\Sigma}{2\sqrt{\left(\frac{B}{2}\right)^2+\Sigma}}\,,
\end{equation}
and with careful treatment of all summations, one obtains
\begin{multline}
e^{-i\Gamma_\mathrm{N} \delta t / 2\hbar} =\\
\left( \begin{array}{cccccccccc}
T_{\cos} - \frac{B T_{\sin}}{2} & 
-(N_1 +1) A_1 T_{\sin} &
-(N_2 +1) A_2 T_{\sin} &
\cdots &
-(N_M +1) A_M T_{\sin} &
-N_1 A_1^\ast T_{\sin} &
-N_2 A_2^\ast T_{\sin} &
\cdots &
-N_M A_M^\ast T_{\sin}
\\ 
-A_1^\ast T_{\sin} &
A_1^\ast (N_1+1) A_1 T_{\rm mix} &
A_1^\ast (N_2+1) A_2 T_{\rm mix} &
\cdots &
A_1^\ast (N_M+1) A_M T_{\rm mix} &
A_1^\ast N_1     A_1^\ast T_{\rm mix} &
A_1^\ast N_2     A_2^\ast T_{\rm mix} &
\cdots &
A_1^\ast N_M     A_M^\ast T_{\rm mix} 
\\ 
-A_2^\ast T_{\sin} &
A_2^\ast (N_1+1) A_1 T_{\rm mix} &
A_2^\ast (N_2+1) A_2 T_{\rm mix} &
\cdots &
A_2^\ast (N_M+1) A_M T_{\rm mix} &
A_2^\ast N_1     A_1^\ast T_{\rm mix} &
A_2^\ast N_2     A_2^\ast T_{\rm mix} &
\cdots &
A_2^\ast N_M     A_M^\ast T_{\rm mix} 
\\ 
\vdots & \vdots &\vdots & \ddots & \vdots & \vdots & \vdots & \cdots & \vdots 
\\ 
-A_M^\ast T_{\sin} &
A_M^\ast (N_1+1) A_1 T_{\rm mix} &
A_M^\ast (N_2+1) A_2 T_{\rm mix} &
\cdots &
A_M^\ast (N_M+1) A_M T_{\rm mix} &
A_M^\ast N_1     A_1^\ast T_{\rm mix} &
A_M^\ast N_2     A_2^\ast T_{\rm mix} &
\cdots &
A_M^\ast N_M     A_M^\ast T_{\rm mix} 
\\ 
-A_1 T_{\sin} &
A_1 (N_1+1) A_1 T_{\rm mix} &
A_1 (N_2+1) A_2 T_{\rm mix} &
\cdots &
A_1 (N_M+1) A_M T_{\rm mix} &
A_1 N_1     A_1^\ast T_{\rm mix} &
A_1 N_2     A_2^\ast T_{\rm mix} &
\cdots &
A_1 N_M     A_M^\ast T_{\rm mix} 
\\ 
-A_2 T_{\sin} & 
A_2 (N_1+1) A_1 T_{\rm mix} &
A_2 (N_2+1) A_2 T_{\rm mix} &
\cdots &
A_2 (N_M+1) A_M T_{\rm mix} &
A_2 N_1     A_1^\ast T_{\rm mix} &
A_2 N_2     A_2^\ast T_{\rm mix} &
\cdots &
A_2 N_M     A_M^\ast T_{\rm mix} 
\\ 
\vdots & \vdots & \vdots & \vdots & \vdots & \vdots & \vdots & \ddots & \vdots 
\\ 
-A_M T_{\sin} &
A_M (N_1+1) A_1 T_{\rm mix} &
A_M (N_2+1) A_2 T_{\rm mix} &
\cdots &
A_M (N_M+1) A_M T_{\rm mix} &
A_M N_1     A_1^\ast T_{\rm mix} &
A_M N_2     A_2^\ast T_{\rm mix} &
\cdots &
A_M N_M     A_M^\ast T_{\rm mix} 
\\ 
\end{array}\right)\\
+\left(\begin{array}{ccccc}
0      & 0           & 0           & \cdots & 0\\
0      & T_{\rm exp} & 0           & \cdots & 0\\
0      & 0           & T_{\rm exp} & \cdots & 0\\
\vdots & \vdots      & \vdots      & \ddots & \vdots\\
0      & 0           & 0           & \cdots & T_{\rm exp}\\
 \end{array} \right)\,,
\label{eqn:app_exp_matrix}
\end{multline}
where
\begin{align}
T_{\rm exp} &= \exp\left(-iH_\mathrm{N} \frac{\delta t}{2\hbar}\right)\,, \\
T_{\rm cos} &= \cos \left(\left[\left(\frac{B}{2}\right)^2 + \Sigma\right]^{\frac{1}{2}} \frac{\delta t}{2 \hbar} \right) \exp\left(-i\left[H_\mathrm{N}+\frac{B}{2}\right]\frac{\delta t}{2 \hbar}\right) \,,\\
T_{\rm sin} &=i\sin \left(\left[\left(\frac{B}{2}\right)^2 + \Sigma\right]^{\frac{1}{2}} \frac{\delta t}{2 \hbar} \right)
 \exp\left(-i\left[H_\mathrm{N}+\frac{B}{2}\right]\frac{\delta t}{2 \hbar}\right) \left[\left(\frac{B}{2}\right)^2 + \Sigma\right]^{-\frac{1}{2}} \,, \\
T_{\rm mix} &= \frac{T_{\rm cos} + \frac{1}{2}T_{\rm sin} B - T_{\rm exp}}{\Sigma}\,.
\end{align}
Thus, the action of the operator $e^{-i\Gamma_\mathrm{N} \delta t / 2\hbar}$
can be succinctly expressed as $2M+1$ coupled equations:
\begin{align}
\label{eqn:result_first}
\psi(\vx,t+\delta t/2) &= \left(T_{\rm cos}(\vx,t) -\frac{1}{2} T_{\rm sin}(\vx,t) B(\vx,t) \right)\psi(\vx,t) - T_{\rm sin}(\vx,t) \Xi(\vx,t)\,,\\
v^\ast_k(\vx,t+\delta t/2) &= -A_k^\ast(\vx,t) T_{\rm sin}(\vx,t) \psi(\vx,t) + T_{\rm exp}(\vx,t) v_k^\ast(\vx,t) + A_k^\ast (\vx,t) T_{\rm mix}(\vx,t) \Xi(\vx,t) \,,\\
u_k(\vx,t+\delta t/2) &= -A_k(\vx,t) T_{\rm sin}(\vx,t) \psi(\vx,t) + T_{\rm exp}(\vx,t) u_k(\vx,t) + A_k (\vx,t) T_{\rm mix}(\vx,t) \Xi(\vx,t)\,,
\end{align}
where
\begin{equation}
\label{eqn:result_last}
\Xi(\vx,t) = \sum_{l=1}^M (N_l+1)A_l(\vx,t)v_l^\ast(\vx,t) + N_l A_l^\ast(\vx,t) u_l(\vx,t)\,,
\end{equation}
and we have explicitly indicated all space- and time-dependence for clarity. Equations. (\ref{eqn:result_first})--(\ref{eqn:result_last}) appear in the main text as Eqs. (\ref{eqn:psi_evo})--(\ref{eqn:main_xi_def}), and constitute the central result of this appendix.
\end{widetext}

%

\end{document}